%% file: paper.tex
\documentclass[twocolumn, twocolappendix]{aastex63}

\usepackage{amsmath}

\def\cii  {\ensuremath{\text{[C\,\textsc{ii]}}}}
\def\ci  {\ensuremath{\text{[C\,\textsc{i]}}}}
\shorttitle{A closer look at two ultra-luminous QSOs}
\shortauthors{Schindler et al.}
\graphicspath{{./}{}}

\begin{document}

\title{A Closer Look at Two of the Most Luminous Quasars in the Universe}

\correspondingauthor{Jan-Torge Schindler}
\email{schindler@mpia.de}

\author[0000-0002-4544-8242]{Jan-Torge Schindler}
\affiliation{Max Planck Institut f\"ur Astronomie, K\"onigstuhl 17, D-69117, Heidelberg, Germany}

\author[0000-0003-3310-0131]{Xiaohui Fan}
\affiliation{Steward Observatory, University of Arizona, 933 North Cherry Avenue, Tucson, AZ 85721, USA} 

\author[0000-0001-8695-825X]{Mladen Novak}
\affiliation{Max Planck Institut f\"ur Astronomie, K\"onigstuhl 17, D-69117, Heidelberg, Germany}

\author[0000-0001-9024-8322]{Bram Venemans}
\affiliation{Max Planck Institut f\"ur Astronomie, K\"onigstuhl 17, D-69117, Heidelberg, Germany}

\author[0000-0003-4793-7880]{Fabian Walter}
\affiliation{Max Planck Institut f\"ur Astronomie, K\"onigstuhl 17, D-69117, Heidelberg, Germany}

\author[0000-0002-7633-431X]{Feige Wang}
\altaffiliation{Hubble Fellow}
\affiliation{Steward Observatory, University of Arizona, 933 North Cherry Avenue, Tucson, AZ 85721, USA} 

\author[0000-0001-5287-4242]{Jinyi Yang}
\affiliation{Steward Observatory, University of Arizona, 933 North Cherry Avenue, Tucson, AZ 85721, USA} 

\author[0000-0002-5367-8021]{Minghao Yue}
\affiliation{Steward Observatory, University of Arizona, 933 North Cherry Avenue, Tucson, AZ 85721, USA} 

\author[0000-0002-2931-7824]{Eduardo Ba{\~n}ados} 
\affiliation{Max Planck Institut f\"ur Astronomie, K\"onigstuhl 17, D-69117, Heidelberg, Germany}

\author[0000-0003-4955-5632]{Yun-Hsin Huang}
\affiliation{Steward Observatory, University of Arizona, 933 North Cherry Avenue, Tucson, AZ 85721, USA}



\begin{abstract}

Ultra-luminous quasars ($M_{1450} \leq -29$) provide us with a rare view into the nature of the most massive and most rapidly accreting supermassive black holes (SMBHs). Following the discovery of two of these extreme sources, J0341${+}$1720 ($M_{1450}=-29.56$, $z=3.71$) and J2125${-}$1719 ($M_{1450}=-29.39$, $z=3.90$), in the Extremely Luminous Quasar Survey (ELQS) and its extension to the Pan-STARRS\,1 footprint (PS-ELQS), we herein present an analysis of their rest-frame UV to optical spectroscopy. Both quasars harbor very massive SMBHs with 
$M_{\rm{BH}}=6.73_{-0.83}^{+0.75}\times10^{9}\,M_{\odot}$ and $M_{\rm{BH}}=5.45_{-0.55}^{+0.60}\times10^{9}\,M_{\odot}$, respectively, showing evidence of accretion above the Eddington limit ($L_{\rm{bol}}/L_{\rm{Edd}}=2.74_{-0.27}^{+0.39}$ and $L_{\rm{bol}}/L_{\rm{Edd}}=3.01_{-0.30}^{+0.34}$).
NOEMA 3\,millimeter observations of J0341${+}$1720 reveal a highly star-forming ($\rm{SFR}\approx1500\,M_{\odot}\,\rm{yr}^{-1}$), ultra-luminous infrared galaxy ($L_{\rm{TIR}}\approx1.0\times10^{13}\,L_{\odot}$) host, which, based on an estimate of its dynamical mass, is only ${\sim}30$ times more massive than the SMBH it harbors at its center.
As examples of luminous super-Eddington accretion, these two quasars provide support for theories, which explain the existence of billion solar mass SMBHs ${\sim}700$ million years after the Big Bang by moderate super-Eddington growth from standard SMBH seeds.

\end{abstract}

\keywords{galaxies: nuclei - quasars: general - quasars: supermassive black holes - quasars: emission lines}

\section{Introduction} \label{sec:intro}

Ultra-luminous quasars host the most massive and/or the most rapidly accreting SMBHs making them prime targets in the search for SMBHs accreting above the Eddington limit. Additionally, their analysis offers insight into a range of science questions, often related to BH formation and evolution:
(i) Their discovery places strong constraints on the bright end of the quasar luminosity function \citep{Schindler2019a}.
(ii) The bright background emission of ultra-luminous quasars facilitates studies of the metal-enrichment in the intervening intergalactic medium \citep{Simcoe2004} and foreground galaxies \citep{RyanWeber2009, Simcoe2011}. 
(iii) Their BH mass estimates provide important constraints on the proposed maximum-mass limit \citep{Inayoshi2016, King2016} of SMBHs, on SMBH seeds, and early BH mass growth. 
(iv) They allow to probe the relation between SMBHs and the formation of the most massive galaxies.

\begin{figure}
    \centering
    \includegraphics[width=0.5\textwidth]{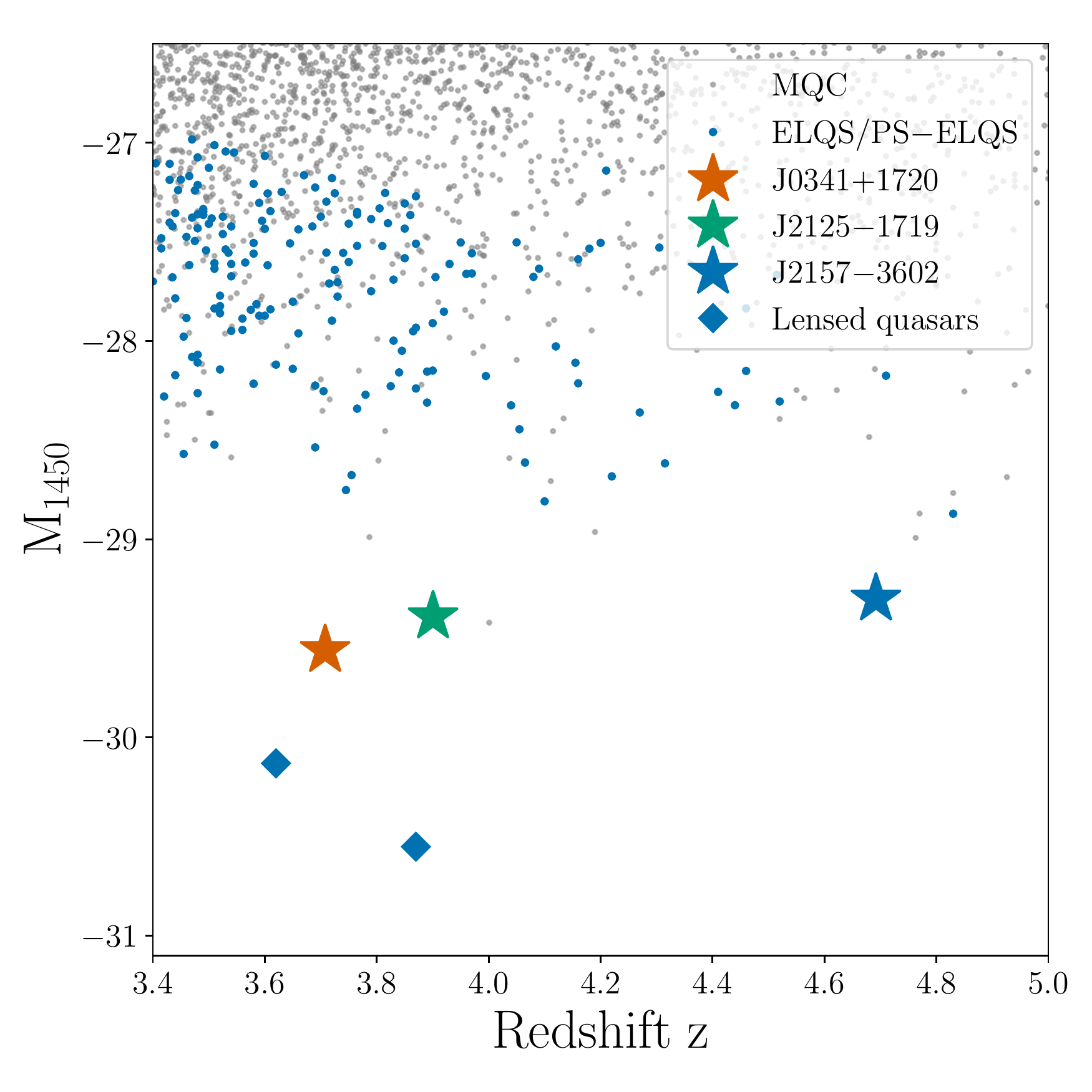}
    \caption{Quasars in the plane of absolute magnitude at 1450\AA, $M_{1450}$, and redshift, $z$. We highlight the two quasars of this study, J0341${+}$1720 and J2125-1710, as orange and green stars, respectively. We further display J2157{-}3602, an ultra-luminous quasar at $z=4.692$ \citep{Wolf2018, Onken2020} as the blue star and two prominent lensed quasars as blue diamonds. Grey dots in the background show the quasar distribution based on the Million Quasar Catalogue \citep[MQC][v5.7b]{Flesch2015}. We calculate the absolute magnitude, $M_{1450}$, for all sources based on the PS1 i-band photometry as in \citep{Schindler2019b}. Quasars discovered as part of the ELQS and PS-ELQS surveys are highlighted in blue.}
    \label{fig:zM1450}
\end{figure}

In part driven by recent efforts to find extremely luminous quasars in the southern hemisphere \citep{Calderone2019,Schindler2019b,Boutsia2020, Wolf2020}, past years have seen the discovery of several ultra-luminous sources \citep{WangFeige2015, Wu2015, Wolf2018, Schindler2019a, Schindler2019b, Jeram2020} at $z\approx2.5-6.5$. In this paper we present follow-up observations on two of them.
J0341${+}$1720 was discovered as part of the Extremely Luminous Quasar Survey (ELQS) survey \citep{Schindler2017} and published in \citet{Schindler2019a} with a redshift of $z=3.69$ determined from the optical discovery spectrum. J2125${-}$1719 was later discovered during an extension of the ELQS to the Panoramic Survey Telescope and Rapid Response System  \citep[Pan-STARRS\,1,][]{Kaiser2002, Kaiser2010}  $3\pi$ survey \citep[PS1,][]{Chambers2016} footprint \citep[PS-ELQS,][]{Schindler2019b} with a discovery redshift of $z=3.90$. 
Table\,\ref{table:sample} provides an overview over the general properties of both quasars, their redshifts, absolute magnitudes at a rest-frame wavelength of $1450$\AA, coordinates and discovery photometry. 
J0341${+}$1720 and J2125${-}$1719 are ultra-luminous quasars with $M_{1450}=-29.46$ and $-29.35$, respectively, as derived from their i-band magnitudes (see Figure\,\ref{fig:zM1450}). In this paper we report on the results from follow-up campaigns providing a much more detailed look into the nature of these two systems. 

We present the near-infrared and optical spectroscopy of J0341${+}$1720 and J2125${-}$1719 in Section\,\ref{sec:nearir_opt}, including the model fitting, analysis and results. Section\,\ref{sec:J0341_mm} describes the NOEMA 3\,mm observations of J0341${+}$1720 and their analysis. In Section\,\ref{sec:discussion} we discuss the possibility of these quasars being lensed, evidence for accretion beyond the Eddington limit, and put the observations of J0341${+}$1720 in context with SMBH galaxy co-evolution.
A brief summary is given in Section\,\ref{sec:conclusion}.
We report all magnitudes in the AB photometric system, unless otherwise noted. For cosmological calculations we have adopted a standard flat $\Lambda$CDM cosmology with H$_0=70\,\rm{km}\,\rm{s}^{-1}\,\rm{Mpc}^{-1}$, $\Omega_{\rm M}=0.3$, and \mbox{$\Omega_\Lambda=0.7$}.

\begin{deluxetable}{ccc}
\tablecaption{General properties of the two quasars\label{table:sample}}
\tablehead{\colhead{Property} &\colhead{J0341${+}$1720} &\colhead{J2125${-}$1719}} 
\startdata 
Discovery reference & ELQS & PS-ELQS \\
Discovery redshift & 3.69 & 3.90 \\
$M_{1450}$ (mag, i-band)\tablenotemark{a} & -29.46 & -29.35\\
\hline
Redshift (this work) & $3.7078\pm0.0005$ & $3.9011\pm0.0003$ \\ 
Redshift method & CO(4-3) & O[III]5007\AA \\
\hline
\multicolumn{3}{c}{--------------- Optical survey ---------------}  \\
Survey & SDSS & PS1 \\
R.A. (degrees) & 55.46319 & 321.42069 \\
$\sigma_{\rm{R.A.}}$ (degrees) & $0.00180$ & $0.00533$\\
Decl. (degrees) & 17.34715 & -17.33095\\
$\sigma_{\rm{Decl.}}$ (degrees) & $0.00184$ & $0.00327$\\
u-band (mag) & $23.036\pm0.779$ & \dots \\
g-band (mag)& $17.394\pm0.014$ & $17.710\pm0.007$ \\
r-band (mag)& $16.383\pm0.020$ & $16.504\pm0.008$ \\
i-band (mag)& $16.193\pm0.014$ & $16.422\pm0.003$ \\
z-band (mag)& $16.085\pm0.014$ & $16.419\pm0.008$ \\
y-band (mag)& \dots & $16.287\pm0.010$ \\
\multicolumn{3}{c}{--------------- 2MASS ---------------}  \\
J-band (mag)& $15.94\pm0.05$ & $16.13\pm0.6$ \\
H-band (mag)& $15.82\pm0.05$ & $16.14\pm0.07$ \\
K-band (mag)& $15.62\pm0.06$ & $16.01\pm0.07$ \\
\multicolumn{3}{c}{-------------- AllWISE ---------------}  \\
W1 (mag) & $15.61\pm0.02$ & $15.92\pm0.03$ \\
W2 (mag) & $15.66\pm0.02$ & $16.31\pm0.03$ \\
W3 (mag) & $14.10\pm0.03$ & $15.56\pm0.09$ \\
W4 (mag) & $13.34\pm0.08$ & $14.96\pm0.33$ \\
\multicolumn{3}{c}{--------------- Gaia ---------------}  \\
R.A. (degrees) & $55.46321$ & $321.42069$\\
$\sigma_{\rm{R.A.}}$ (mas) & $0.07402$ & $0.09780$\\
Decl. (degrees) & 17.34716 & -17.33095\\
$\sigma_{\rm{Decl.}}$ (mas) & $0.05345$ & $0.08838$\\
G (mag)\tablenotemark{b} & 16.71 & 16.87\\
BP (mag)\tablenotemark{b} & 17.37 & 17.46 \\
RP (mag)\tablenotemark{b} & 15.95 & 16.16\\
\enddata 
\tablenotetext{a}{We include the values for $M_{1450}$ (mag, i-band) here for comparison with quasars in Figure\,\ref{fig:zM1450}. In Table\,\ref{tab:fitproperties} we provide the updated values derived from the spectral fits.}
\tablenotetext{b}{Gaia magnitudes are in the Vega system. The Gaia DR2 does not provide an error for this quantity as the error distribution is only symmetric in flux space.}
\end{deluxetable} 

\section{Optical and near-infrared spectroscopy}\label{sec:nearir_opt}

\begin{figure*}
    \centering
    \includegraphics[width=\textwidth]{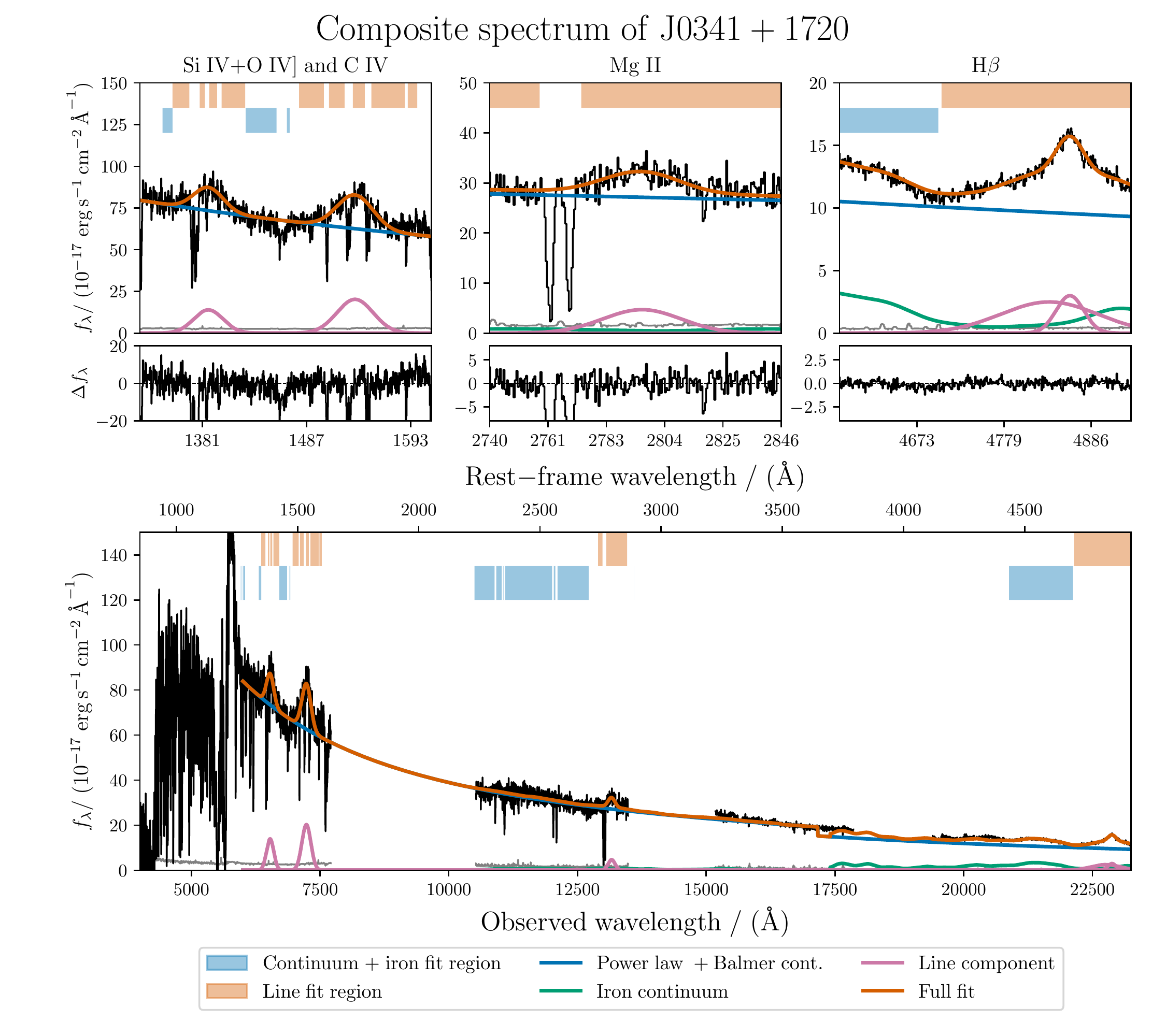}
    \caption{Optical to near-infrared composite spectrum of J0341${+}$1720 including the best-fit model from our analysis. The lower panel shows the full composite spectrum, while the upper panels display the emission line fits around the \ion{Si}{4}{+}O~IV] and \ion{C}{4}, \ion{Mg}{2}, and H$\beta$ lines in more detail. The best-fit model is highlighted as the solid orange line. Solid blue, green and pink lines show the individual power law and Balmer continuum, iron continuum and line component models. Flux uncertainties are given by the grey line. Light blue and orange bars at the top of each panel highlight which wavelength regions were included in the continuum (power law + Balmer continuum \& iron FeII continuum) and emission line fits. The presented best-fit model uses the \citet{Tsuzuki2006} iron template in the \ion{Mg}{2} region and the \citet{Boroson1992a} iron template around H$\beta$.}
    \label{fig:specJ0341}
\end{figure*}

\subsection{J0341${+}$1720 observations}\label{sec:J0341_obs_data}

The two optical spectra of J0341${+}$1720 were taken on 2016 December 19 using the VATTSpec spectrograph on the Vatican Advanced Technology Telescope (VATT). We used the  $300\,$g/mm grating in first order blazed at $5000\,\text{\AA}$. The spectra have a resolution of $R\sim1000$ ($1\farcs5$ slit) and a coverage of $\sim4000\,\text{\AA}$ around our chosen central wavelength of $\sim5775\,\text{\AA}$. Each spectrum was exposed for $900\,\rm{s}$. Observations are carried out with the slit aligned with the parallactic angle to minimize slit losses. 

We reduced the optical spectra using the standard long slit reducion routines within the IRAF software \citep{Tody1986, Tody1993}. One dimensional spectra were extracted with the apall routine using the built-in cosmic-ray removal and optimal extraction \citep{Horne1986}. We calibrated the wavelength using the internal VATTSpec HgAr lamp. Relative flux calibration was done using a standard flux calibrator. The spectra have not been corrected for telluric absorption. 


The near-infrared follow-up spectroscopy was taken on 2018 November 24 using the LUCI\,1 and LUCI\,2 instruments in binocular mode on the Large Binocular Telescope (LBT). We used the G200, $200\,$g/mm grating, on LUCI\,1 in the HKspec filter band covering the wavelength range of $13200\,\text{\AA}$ to $24000\,\text{\AA}$ with a resolution of $R\approx630$ to $870$ ($1\farcs5$ slit). The simultaneously executed observations with LUCI\,2 used the G200, $200\,$g/mm grating, in the zJspec filter band covering a wavelength range of $9000\,\text{\AA}$ to $12500\,\text{\AA}$ with a resolution of $R\approx700$ to $800$ ($1\farcs5$ slit). 
Four exposures of $150\,\rm{s}$ each were taken in a standard ABBA dithering pattern. 


The near-infrared spectra were reduced with the open source ``Python Spectroscopic Data Reduction Pipeline'', \texttt{PypeIt}\footnote{https://github.com/pypeit/PypeIt} \citep{PypeitProchaska2019, PypeitProchaska2020}. 
The pipeline automatically traces the long-slit spectra and corrects for detector illumination. Skylines are subtracted by difference imaging on the dithered AB pairs and with a 2D BSpline fitting procedure. 
The 1D spectra are extracted using the optimal spectrum extraction technique \citep{Horne1986} along automatically identified traces.  
Relative flux calibration is done using the standard star HD24000 observed during the same night. The four individual exposures are then co-added and corrected for telluric absorption. The pipeline uses a large grid of telluric models produced from the Line-By-Line Radiative Transfer Model \citep[LBLRTM4,][]{Clough2005, Gullikson2014} to find the best-fit model, which corrects the absorbed quasar spectrum up to a best-fit PCA model \citep{Davies2018d} of it.

Due to the changing and non-optimal weather conditions absolute flux calibration using the standard stars is not reliable. After co-adding the optical spectra we have flux calibrated them using the SDSS r-band magnitude of the quasar. 
The zJband and HKband spectra were similarly flux calibrated using the J-band and K-band photometry of 2MASS. 
All spectra have been extinction corrected\footnote{We used the python package \texttt{extinction} \citep{python_extinction} to calculate the extinction correction.} using the extinction law of \citet{Fitzpatrick2007} with $R_V=3.1$ and $A_V=0.45$ \citep{Schlafly2011}.

\subsection{J2125${-}$1719 observations}\label{sec:J2125_obs_data}


The optical discovery spectra of J2125${-}$1719 were taken on 2017 October 7 and 10 with the Goodman High Throughput Spectrograph (Goodman HTS) on the Southern Astrophysical Research (SOAR) Telescope ($4.1\,\rm{m}$). 
Observations using the 400\,g/mm grating with central wavelengths of $6000\,\text{\AA}$ and $7300\,\text{\AA}$ resulted in spectra with a wavelength coverage of $\sim4000-8000\,\text{\AA}$ (GG-385 blocking filter) and $\sim5300-9300\,\text{\AA}$ (GG-495 blocking filter), respectively. All observations used the red camera in 2x2 spectral binning mode. The $\sim4000-8000\,\text{\AA}$ ``blue'' spectrum was observed for $600\,\rm{s}$ with the $1\farcs2$ slit resulting in a resolution of $R\approx690$. For the $\sim5300-9300\,\text{\AA}$ "red" spectrum we exposed for $900\,\rm{s}$ using the $1\farcs0$ slit, which provides a slightly better resolution of $R\approx830$. 
The spectra were reduced with IRAF using the same methods as for the optical spectra in Section\,\ref{sec:J0341_obs_data}.

The near-infrared follow-up spectroscopy was carried out with the Folded-port InfraRed Echellete spectrograph \citep[FIRE,][]{Simcoe2013} at Magellan Baade. We have taken a FIRE Echelle spectrum using the $0\farcs75$ slit over the nominal spectral range of $\sim8250-25200\,$\AA\ with a resolution of $R\approx6000$ on 2018 September 25.
The near-infrared FIRE spectrum was also reduced using \textit{PypeIt} as described for J0341${+}$1720 in Section\,\ref{sec:J0341_obs_data}.
The optical and near-infrared spectra were also extinction corrected using the extinction law of \citet{Fitzpatrick2007} with $R_V=3.1$ and $A_V=0.16$ \citep{Schlafly2011}.

To generate the composite spectrum the near-infrared spectrum was flux calibrated to the 2MASS K-band magnitude. We scaled the "red" SOAR spectrum to the near-infrared spectrum in the region of $8500-9000\,\text{\AA}$ and joined them at $8500\,\text{\AA}$. We further added the "blue" SOAR spectrum to the composite blueward of $6500\,\text{\AA}$, scaling the "blue" SOAR spectrum flux to match the composite in the $6000-7000\,\text{\AA}$ overlap region.


\subsection{Modeling of the optical and near-infrared spectroscopy} \label{sec:spec_modeling}

\begin{figure*}
    \centering
    \includegraphics[width=\textwidth]{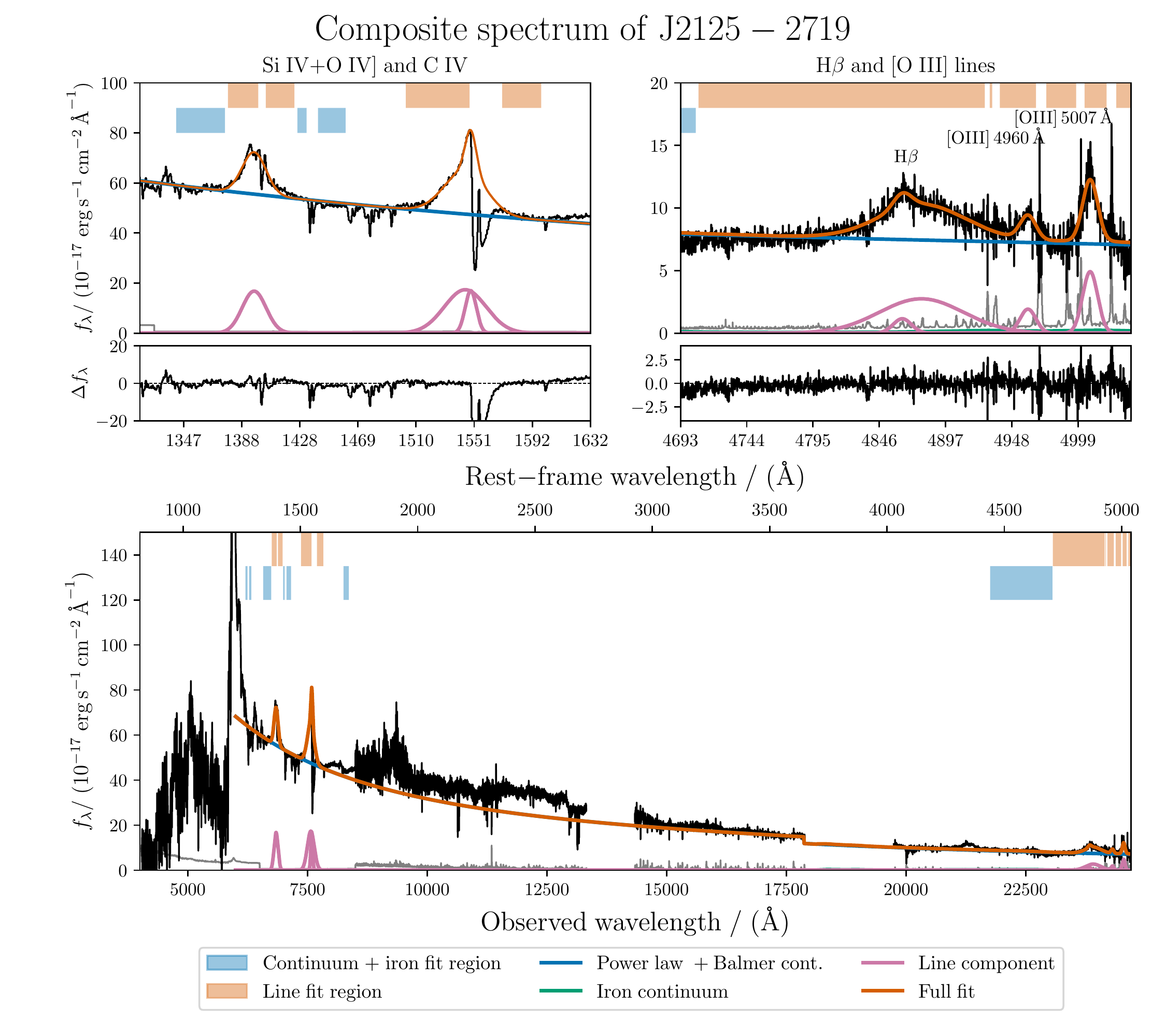}
    \caption{Optical to near-infrared composite spectrum of J2125${-}$1719 including the best-fit model from our analysis. The lower panel shows the full composite spectrum, while the upper panels display the emission line fits of \ion{Si}{4} \& \ion{C}{4} and H$\beta$ \& [O~III] in more detail. For the description of the lines and symbols, see Figure\,\ref{fig:specJ0341}.
    }
    \label{fig:specJ2125}
\end{figure*}
A general description of our fitting procedure, detailing the continuum models, line models and basic wavelength regions included in the fit, is given in Appendix\,\ref{app:fitmethod}.  
Figures\,\ref{fig:specJ0341} and \ref{fig:specJ2125} show the full composite spectra and their best-fit model (solid orange line) for J0341${+}$1720 and J2125${-}$1719, respectively. The bottom panel in each figure shows the full wavelength coverage, whereas the top panels provide a detailed view on the emission lines of interest.
We masked out regions severely affected by telluric absorption or reduction artifacts at the edges. This reduces the nominal coverage of the LUCI\,1/2 spectra of J0341${+}$1720 in Figure\,\ref{fig:specJ0341}.
In the following we describe details on the spectral modeling specific to each quasar and exceeding the general description in Appendix\,\ref{app:fitmethod}. 

We have fit J0341${+}$1720 twice using both the \citet{Tsuzuki2006} and the \citet{Vestergaard2001} iron templates. The only property we adopt from the fit with the \citet{Vestergaard2001} template is the FWHM of \ion{Mg}{2} to estimate the \ion{Mg}{2}-based black hole mass using the  \citet{Vestergaard2009} relation.
To model the composite spectrum of J0341${+}$1720 we approximate each of the \ion{Si}{4}{+}O~IV], \ion{C}{4} and \ion{Mg}{2} lines with one Gaussian component. To accurately describe the H$\beta$ line we use two Gaussian components. 
The \ion{Si}{4} line appears to be broad and significantly blueshifted. We thus reduce the blue-ward $\lambda_{\rm{rest}} = 1340{-}1375\,\text{\AA}$ continuum window, to fit the line in a larger rest-frame wavelength window of $\lambda_{\rm{rest}} = 1350{-}1425\,\text{\AA}$. For the \ion{C}{4}, \ion{Mg}{2}, and the H$\beta$ and [O~III] lines we use the wavelength ranges of $\lambda_{\rm{rest}} = 1480{-}1600\,\text{\AA}$, $\lambda_{\rm{rest}} = 2740{-}2860\,\text{\AA}$, and $\lambda_{\rm{rest}} = 4700{-}5050\,\text{\AA}$, respectively. 
We interactively mask out regions affected by strong absorption features, leading to the emission line windows as seen in Figure\,\ref{fig:specJ0341} (light orange bars).

In the case of J2125${-}$1719 we only incorporate the iron template of \citet{Boroson1992a} as we do not fit the \ion{Mg}{2} region. 
One Gaussian component is used to approximate the \ion{Si}{4}{+}O~IV], [O~III] $\lambda4960\,\text{\AA}$, and [O~III] $\lambda5007\,\text{\AA}$ emission lines, while we use two Gaussian components for the \ion{C}{4} and H$\beta$ lines.
We fit the \ion{Si}{4}{+}O~IV], \ion{C}{4}, and the H$\beta$ \& [O~III] emission lines in the wavelength ranges of 
$\lambda_{\rm{rest}} = 1375{-}1425\,\text{\AA}$,
$\lambda_{\rm{rest}} = 1500{-}1600\,\text{\AA}$, and
$\lambda_{\rm{rest}} = 4700{-}5050\,\text{\AA}$, respectively. 
We have masked out the wavelength region of $\lambda = 7586{-}7679\,\text{\AA}$ from the \ion{C}{4} emission line fit as it is strongly affected by telluric absorption features not corrected in the reduction process. In addition, we mask out regions affected by narrow absorption lines as well as residuals from strong sky lines (see Figure\,\ref{fig:specJ2125}).

{\input{prop_table}}

{\input{line_table}}

\subsection{Results of the optical and near-infrared spectroscopy}

Tables\,\ref{tab:fitproperties} and \ref{tab:line_measurements} summarize the measurements of the continuum and line properties from our fit analysis. The tables provide the median with the associated $13.6$ and $86.4$ percentile uncertainties of our re-sampled posterior distribution (see Appendix\,\ref{app:fitmethod}). 

We would like to caution the reader that these results can be sensitive to assumptions made when modeling the spectra, such as the choice of the continuum windows. In the case of high signal-to-noise ratio data, the impact of these assumptions can be larger than the statistical uncertainties from the fitting. Therefore, we have been fully transparent about all assumptions we made, as detailed in Section\,\ref{sec:spec_modeling} and Appendix\,\ref{app:fitmethod}.

%
Our models provide a global fit to the full wavelength range of both spectra. In these cases the continuum fit can over- or underestimate specific regions of the spectrum. As an example, we quantify differences in the continuum properties by directly measuring the absolute magnitude at $1450$\AA\, from the spectrum. 
We measure $M_{1450}=-29.529\pm0.043$ and $M_{1450}=-29.422\pm0.012$ for J0341${+}$1720 and J2125${-}$1719, respectively, showing differences of $|\Delta M_{1450}|=0.04$ to the fit measurements, about a factor of $\sim10$ larger than the pure statistical uncertainties given in Table\,\ref{tab:fitproperties}.

Line redshifts are measured from the peak flux wavelength of the full emission line models. This means that line models made from multiple components are summed before the peak flux wavelength is determined. 
We derive velocity shifts from the line redshifts with respect to the systemic quasar redshift using \textit{linetools} \citep{linetools2016} including relativistic corrections.
For J0341${+}$1720 we adopt the CO(4-3) line redshift, $z=3.7078$ (see Section\,\ref{sec:J0341_mm}), as the systemic redshift for J0341${+}$1720. In the case of J2125${-}$1719 the redshift of the [O~III]$_{5007\text{\AA}}$ line serves as our best estimate for its systemic redshift.

For each line we calculate the FWHM, equivalent width (EW), integrated flux and integrated luminosity from the full line model, including all Gaussian components. 
The spectral resolution (R) can broaden the lines and therefore we correct each line FWHM by:
\begin{equation}
    \rm{FWHM} = \sqrt{\rm{FWHM}_{\rm{obs}}^2 - \rm{FWHM}_{\rm{R}}^2} \ .
\end{equation}

In the case of J0341${+}$1720 we determine the integrated flux and the luminosity of the \ion{Fe}{2} pseudo-continuum in the wavelength range of $2200-3090\,\text{\AA}$, to construct the \ion{Fe}{2}/\ion{Mg}{2} flux ratio, a measure for BLR iron enrichment in high-redshift quasars \citep[][]{Dietrich2003c}. This wavelength region was chosen to be comparable with other studies in the literature \citep[e.g.,][]{Dietrich2003c, Maiolino2003, Kurk2007, DeRosa2011, Mazzucchelli2017}. We calculate a value of $F_{\rm{FeII}}/F_{\rm{MgII}} = 2.78^{+0.79}_{-0.65}$, which is on the lower end of the distribution compared to quasars at $z>3$ \citep[for a comparison, see][their Figure\,5, filled symbols]{Onoue2020}. The inset of the \ion{Mg}{2} line in Figure\,\ref{fig:specJ0341} highlights the small iron contribution around \ion{Mg}{2}. The iron flux around the H$\beta$ line is much stronger, highlighting how the iron contribution can vary throughout the spectrum. 
%

\subsubsection{Bolometric luminosity, black hole mass and Eddington luminosity ratio} \label{sec:derived_prop_a}

The spectral properties measured from the model fits allow us to derive the bolometric luminosity, the black hole mass and the Eddington luminosity ratio. 
We calculate the bolometric luminosity following \citet{Shen2011}:
\begin{equation}
    L_{\rm{bol}} = 5.15\cdot \lambda L_{\lambda, 3000}
\end{equation}
We estimate black hole masses assuming that the line-emitting gas of the \ion{Mg}{2} and H$\beta$ lines is in virial motion around the SMBH. Then the line-of-sight velocity dispersion traces the gravitational potential of the SMBH mass ($M_{\rm{BH}}$). 
Based on reverberation mapping results \citep[e.g.,][]{Onken2004, Peterson2004}, single-epoch virial estimators \citep[e.g.,][]{Vestergaard2006, Vestergaard2009} allow to infer the SMBH mass by measuring the FWHM of broad quasar emission lines and their continuum luminosities. However, these measurements have large systematic uncertainties of $\sim0.4-0.55\,\rm{dex}$ \citep{Vestergaard2009, Shen2013review}.
Single-epoch virial estimators are often written as
\begin{equation}
M_{\rm{BH}} = 10^{zp(x)} \cdot \left[ \frac{\rm{FWHM}}{1000\,\rm{km}\,\rm{s}^{-1}} \right]^2 \left[ \frac{x L_{\lambda, x}}{10^{44}\,\rm{erg}\,\rm{s}^{-1}} \right]^{b} \,M_\odot
\end{equation}
where $zp$, the zero-point BH mass, and the slope $b$ depend on the chosen emission line and continuum luminosity.
We derive black hole masses based on the \ion{Mg}{2}, H$\beta$ and \ion{C}{4} line.
Using the \ion{Mg}{2} line we estimate SMBH masses using the single-epoch virial estimators of \citet[][$zp=6.86$, $b=0.5$, $x=3000\,$\AA]{Vestergaard2009}. 
For the H$\beta$ line we adopt the virial estimator of \citet[][$zp=6.91$, $b=0.5$, $x=5100\,$\AA]{Vestergaard2006}.

Contrary to \ion{Mg}{2} and H$\beta$, the \ion{C}{4} line often shows asymmetric line profiles in addition to large velocity blueshifts \citep[e.g.,][]{Richards2011}. The blueshifts have been shown to correlate with the equivalent width and the FWHM of the line, resulting in biases for black hole masses derived from \ion{C}{4} \citep[e.g.,][]{Shen2013review, Coatman2016, MejiaRestrepo2018a}. A range of studies have thus proposed corrections for these biases \citep[e.g.,][]{Denney2012, Park2013, Runnoe2013c, MeijaRestrepo2016, Coatman2017}, which are not always applicable to all quasars, e.g. depending on their luminosity or Eddington ratio \citep{MejiaRestrepo2018a}.
Still, we decided to derive \ion{C}{4}-based black hole masses as well to compare them with the measurements from the \ion{Mg}{2} and H$\beta$ lines. We adopt  \citet[][$zp=6.66$, $b=0.53$, $x=1350\,$\AA]{Vestergaard2006} and also provide corrected BH masses according to Equations\,4 and 6 of \citet{Coatman2017}. 

We calculate the Eddington luminosity ratio by dividing the bolometric luminosity by the Eddington luminosity:
\begin{equation}
    \lambda_{\text{Edd}} = \frac{L_\text{bol}}{1.26\times10^{38}\,\text{erg}\,\text{s}^{-1} } \frac{M_{\odot}}{M_{\text{BH}}} \ .
\end{equation}

As the absolute magnitude of J0341${+}$1720 already suggests (see Figure\,\ref{fig:zM1450}), this quasar is one of the most luminous in the observable universe with a bolometric luminosity of $L_{\rm{bol}}=2.32(\pm0.01)\times10^{48}\,\rm{erg}\,\rm{s}^{-1}$. 
We list all black hole mass estimates and subsequent Eddington luminosity ratios in Table\,\ref{tab:fitproperties}. 
The black hole mass estimates range from $6.73$ to $33.80\times10^{9}\,M_\odot$. The large values are driven by the \ion{C}{4} line, which shows large \ion{C}{4} blueshifts. Even the correction of \citet{Coatman2017} cannot resolve the tension between the \ion{C}{4} and the H$\beta$ or \ion{Mg}{2} BH mass estimates. Therefore, we regard the \ion{C}{4} measurements as unreliable and only adopt the \ion{Mg}{2} and H$\beta$ black hole masses, $7.47\times10^{9}\,M_\odot$ and $6.73\times10^{9}\,M_\odot$, respectively.
Based on these BH mass measurements we find Eddington luminosity ratios of $2.47$ (\ion{Mg}{2}) and $2.74$ (H$\beta$), suggesting that J0341${+}$1720 is accreting above the Eddington limit. 

With a bolometric luminosity of $L_{\rm{bol}}=2.07\times10^{48}\,\rm{erg}\,\rm{s}^{-1}$, J2125${-}$1719 is also one of the most luminous quasars at $z>3.5$. 
The \ion{C}{4} line has only a very small blueshift and it is not necessary to apply the correction of \citet{Coatman2017} to the BH mass. 
The black hole mass estimates of \ion{C}{4} (VO06) and H$\beta$ (VO06) show good agreement, with values of $4.53$ and $5.45\times10^9\,M_{\odot}$, respectively. The subsequent Eddington luminosity ratios are $\lambda_{\rm{Edd,CIV}}=3.63$ and $\lambda_{\rm{Edd},\rm{H}\beta}=3.01$, i.e. resulting in super-Eddington accretion.

\section{Millimeter observations of J0341${+}$1720}\label{sec:J0341_mm}

To determine the redshift, star formation rate and dynamical mass of the quasar host galaxy of J0341${+}$1720, we observed its CO(4-3) transition at rest-frame $650.3\,\mu\rm{m}$ ($\nu_{\rm{rest}}=461.04\,\rm{GHz}$) using the Northern Extended Millimeter Array (NOEMA). At the redshift of the quasar, $z=3.7078$, the CO(4-3) line shifts to $97.93\,\rm{GHz}$ or $3061.6\,\mu\rm{m}$ observable in Band 1 ($3\,\rm{mm}$). J0341${+}$1720 was observed on 2019 July 1st, 3rd, 5th, and 6th with 9 antennas in configuration D for a total of $9.0\,\rm{hr}$ on--source time. After applying quality cuts to the data we present results on $3.75\,\rm{hr}$ of effective on--source time.
The sources J0322+222 and J0342+147 were observed every 30\,\rm{min} for phase and amplitude calibration. The absolute flux calibration was done using 3C454.3, MWC349, and LKHA101 observed at the beginning of each track. The PolyFix correlator provides a bandwith of $7.7\,\rm{GHz}$ in each of its two sidebands for a total of $15.4\,\rm{GHz}$. We chose a tuning frequency of $99.32\,\rm{GHz}$ to allow for simultaneous detection of the \ci{} line at a rest frequency of $492.16\,\rm{GHz}$ in the upper side band in addition to the targeted CO(4-3) transition.  The data was reduced and analyzed using the Grenoble Image and Line Data Analysis System (GILDAS) software\footnote{http://www.iram.fr/IRAMFR/GILDAS}. 
We re-binned the data to a resolution of $40\,\rm{MHz}$ ($\approx120\,\rm{km}\,\rm{s}^{-1}$), resulting in an average flux (rms) uncertainty of $0.23\,\rm{mJy}\,\rm{beam}^{-1}$. 

\begin{figure*}
    \centering
    \includegraphics[width=0.48\textwidth]{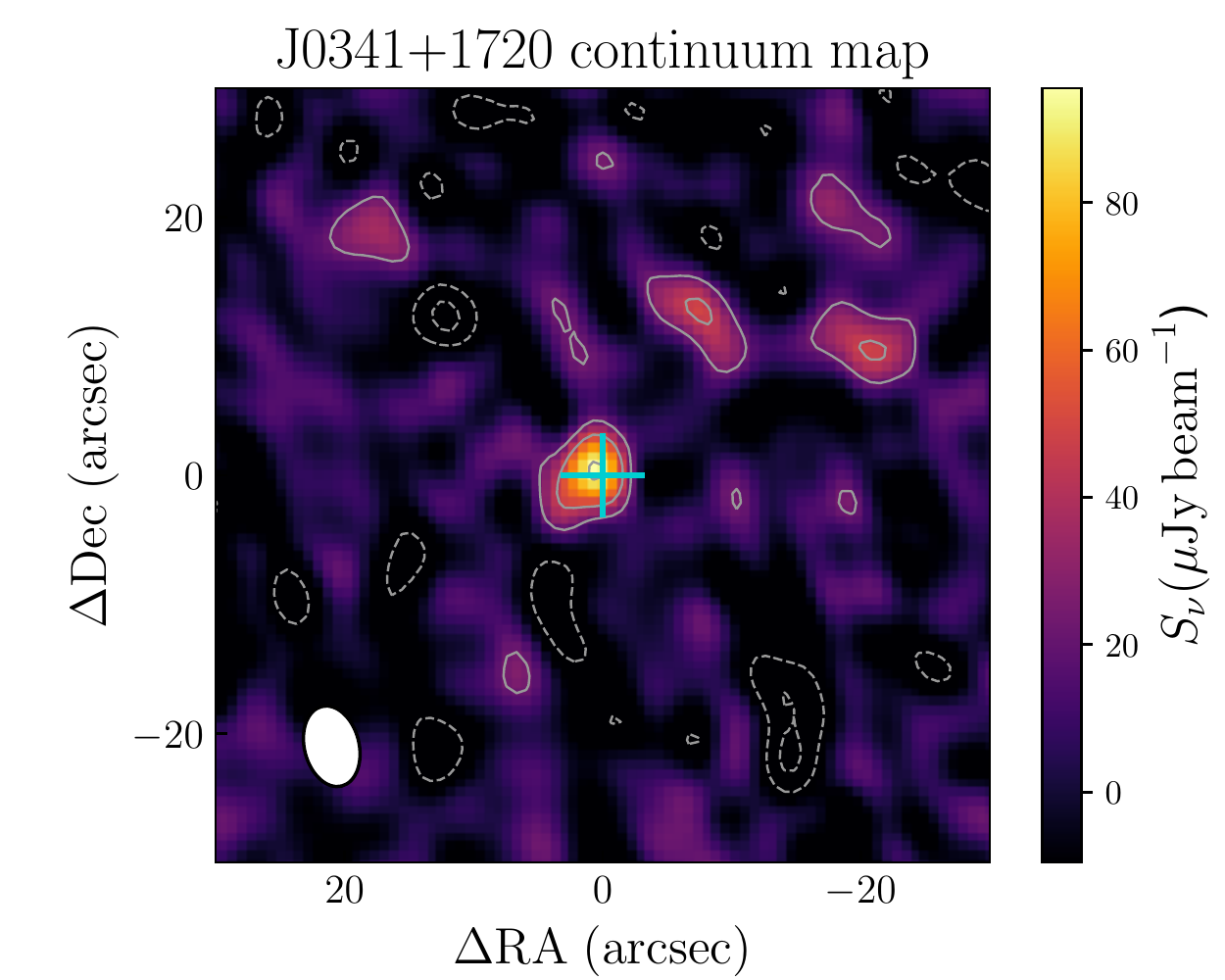}
    \includegraphics[width=0.48\textwidth]{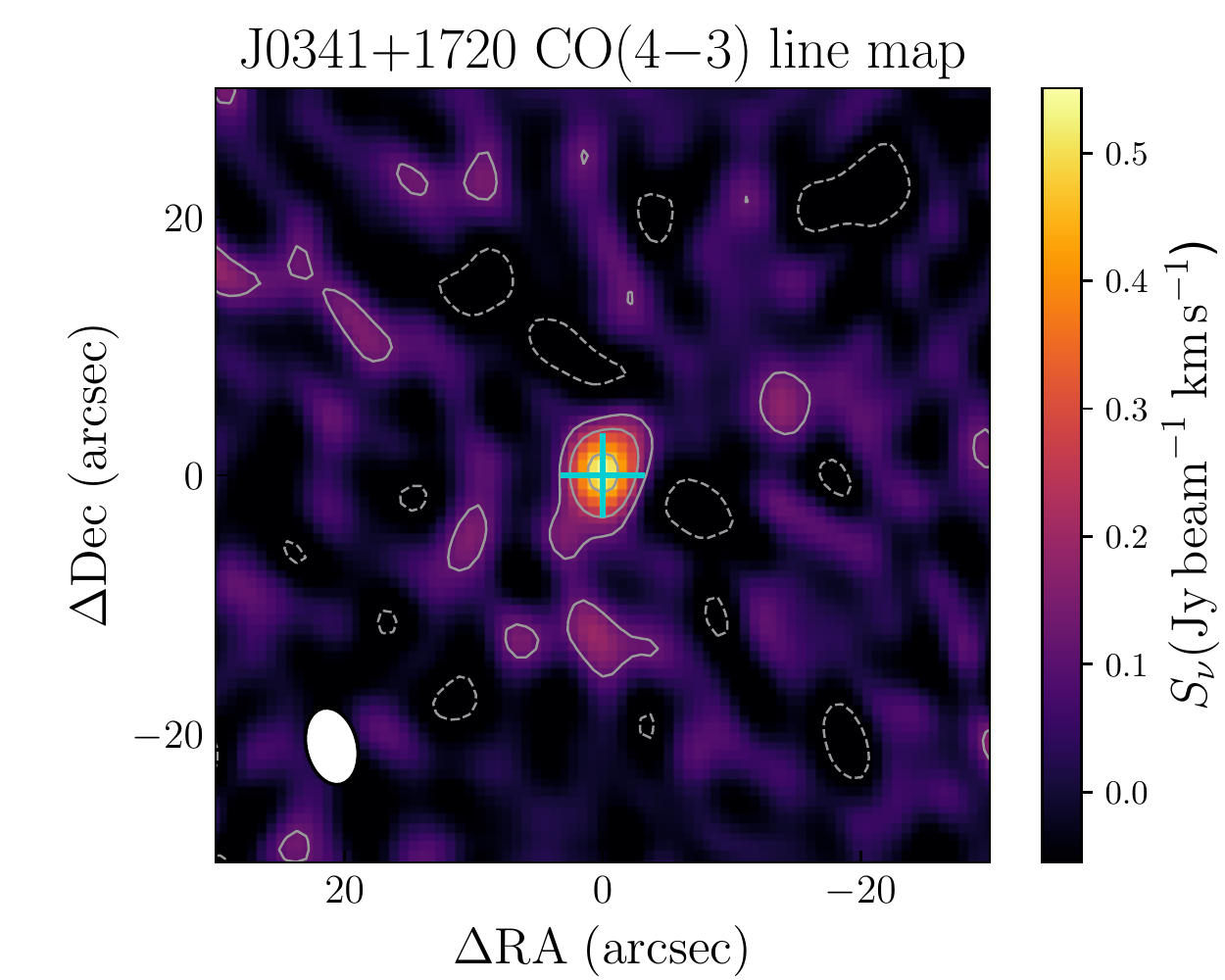}
    \caption{Intensity maps of continuum (left) and CO(4-3) line emission (right) in J0341${+}$1720. Solid and dashed grey contours highlight regions of positive and negative $2,4,8\,\sigma$ and the cyan cross marks the optical position of the quasar.
    \textbf{Left:} Collapsed continuum emission in line-free channels across both side bands ($\sigma=11.6\,\mu\rm{Jy}\,\rm{beam}^{-1}$). The beam FWHM of $6.4\arcsec \times 4.2\arcsec$, shown in the left bottom corner as a white ellipse, corresponds to $\sim45\,\rm{kpc}\times30\,\rm{kpc}$ at the redshift of the quasar. The observations do not resolve the source.
    \textbf{Right:} Continuum subtracted CO(4-3) line emission of J0341${+}$1720 averaged over $\nu=97.6-98.2\,\rm{GHz}$ ($\sigma=0.06\,\rm{Jy}\,\rm{beam}^{-1}\,\rm{km}\, \rm{s}^{-1}$). In the left bottom corner we show the beam FWHM of $6.1\arcsec \times 3.9\arcsec$ as a white ellipse.}
    \label{fig:J0341_maps}
\end{figure*}

Figure\,\ref{fig:J0341_maps} (left panel) shows the continuum map of the joint lower and upper side band, including only line-free channels. The final resolution of the image (natural weighting) is $6.4\arcsec \times 4.2\arcsec$ at a position angle of $14^{\circ}$. 
At the source redshift $1\arcsec$ corresponds to approximately $7\,\rm{kpc}$, which means that the quasar host galaxy is unresolved in our NOEMA observations (e.g., see Venemans et al. 2020, subm.).


We fit a point source in the UV-plane to the continuum map, finding a peak flux of $0.104\,\rm{mJy}\,\rm{beam}^{-1}$ across both side bands with a central frequency of $93.5\,\rm{GHz}$. This results in a continuum detection with a signal-to-noise ratio of S/N$>8$ ($\sigma{=}11.6\,\mu\rm{Jy}\,\rm{beam}^{-1}$) . 
The continuum subtracted collapsed line map is displayed in 
Figure\,\ref{fig:J0341_maps} (right panel), also displaying a significant detection. Natural weighting results in a resolution of $6.1\arcsec \times 3.9\arcsec$ at a position angle of $14^{\circ}$ for the CO(4-3) line.

\subsection{CO(4-3) Line Fit and Estimates of the Host Galaxy's Gas Mass, Dynamical Mass, and Star Formation Rate}

\begin{figure}
    \centering
    \includegraphics[width=0.48\textwidth]{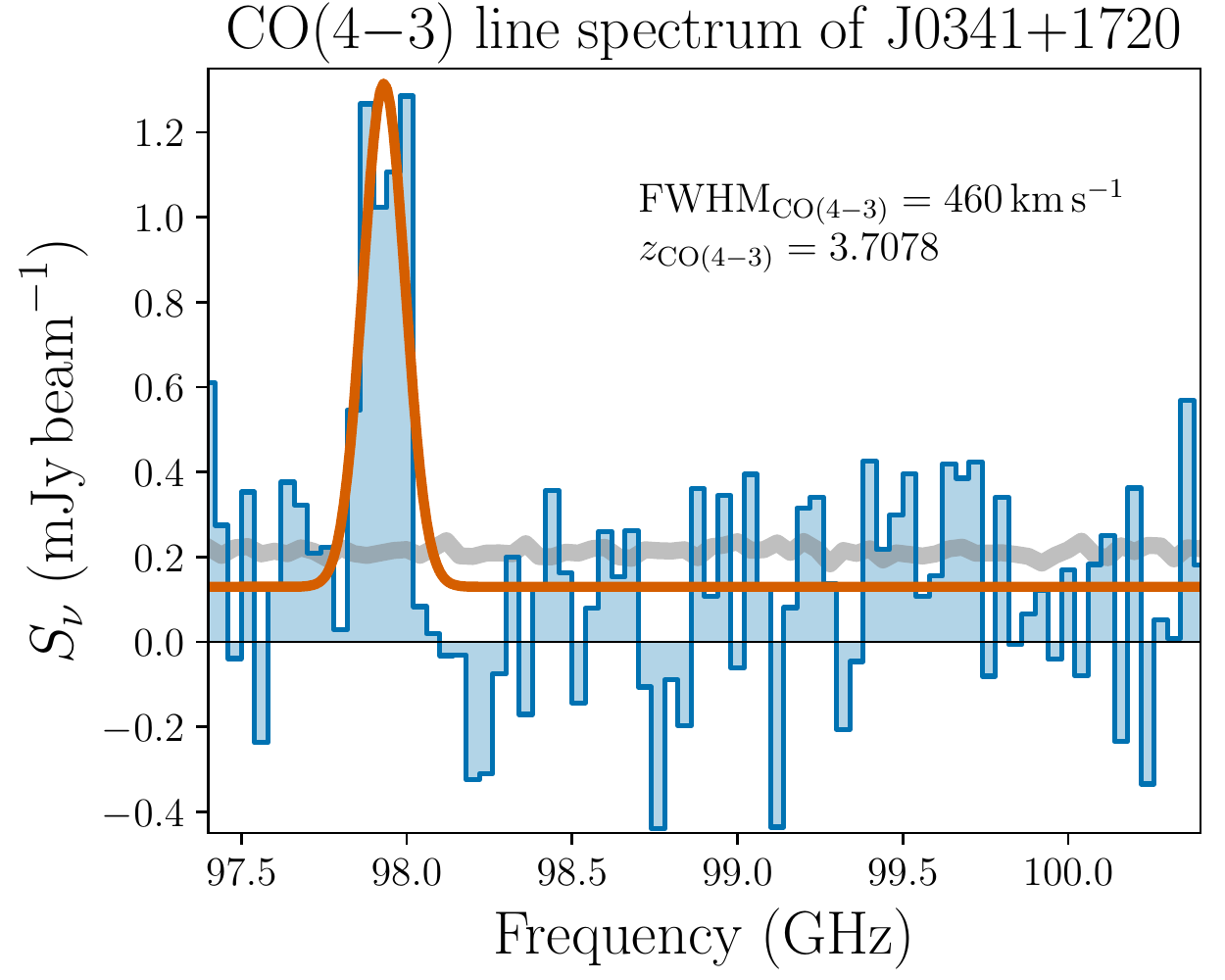}
    \caption{The upper side band spectrum of J0341${+}$1720 at a resolution of $40\,\rm{MHz}$ (blue). The CO(4-3) line is clearly detected at $\nu=97.93\,\rm{GHz}$. We display the flux (rms) uncertainty in grey and over-plot the best-fit model to the spectrum (continuum + line) with the solid orange line.}
    \label{fig:J0341_CO_spec}
\end{figure}

Figure\,\ref{fig:J0341_CO_spec} shows the spectrum in the upper side band with the CO(4-3) line clearly detected around $97.93\,\rm{GHz}$. 
We have chosen a binning of 40\,MHz to ensure that we detect the FWHM of the CO(4-3) at the $\sim3\sigma$ level. At this reduced resolution instrumental effects are negligible compared to the flux uncertainties.
The \ci{} line is not detected, but based on the noise level we can provide an $3\sigma$ upper limit of $\sim0.2\,\rm{Jy}\,\rm{beam}^{-1}\,\rm{km}\,\rm{s}^{-1}$ choosing the same line width as for the CO(4-3) line ($\rm{FWHM}=460\,\rm{km}\,\rm{s}^{-1}$, see below).
A simple fit to the spectrum, shown as the solid orange line, approximates the spectrum with a Gaussian profile for the line and a constant flux value for the continuum. 
Based on our best fit model we measure the line center at
$97.9312\pm0.0095\,\rm{GHz}$, which corresponds to a redshift of $z_{\rm{CO(4-3)}}=3.7078\pm0.0005$ for J0341${+}$1720. The CO(4-3) peak line flux is $1.19\pm0.15\,\rm{mJy}\,\rm{beam}^{-1}$ with a line width of $\sigma_v=63.76\,\rm{MHz}$.
We obtain an integrated line flux of $F_{\rm{CO(4-3)}}=0.58\pm0.12\,\rm{Jy}\,\rm{km}\,\rm{s}^{-1}$ and convert it to a line luminosity following, e.g. \citet{Carilli2013}:
\begin{equation}
\frac{L_{\rm{line}}}{L_\odot} = 1.04\cdot10^{-3} \frac{F_{\rm{line}}}{\rm{Jy}\,\rm{km}\,\rm{s}^{-1}} \frac{\nu_{\rm{obs}}}{\rm{GHz}} \left( \frac{D_{\rm{L}}}{\rm{Mpc}} \right)^{2} \ ,
\end{equation}
where $\nu_{\rm{obs}}$ is the observed frequency of the line and $D_{\rm{L}}$ is the cosmological luminosity distance. We calculate a CO(4-3) line luminosity of $6.3(\pm1.3)\times10^7\,L_\odot$.
We further evaluate the (areal) integrated source brightness temperature \citep[also see][]{Carilli2013} for the CO(4-3) line:
\begin{equation}
\frac{L'_{\rm{line}}}{\rm{K}\,\rm{km}\,\rm{s}^{-1}\,\rm{pc}^2} = 
\frac{3.25\cdot10^7}{\left(1+z\right)^3}
\frac{F_{\rm{line}}}{\rm{Jy}\,\rm{km}\,\rm{s}^{-1}}
\left( \frac{D_{\rm{L}}}{\rm{Mpc}} \cdot \frac{\rm{GHz}}{\nu_{\rm{obs}}}\right)^{2}
\end{equation}
The resulting CO(4-3) integrated source brightness temperature is $L'_{\rm{CO(4-3)}} = 2.0\times10^{10}\,\rm{K}\,\rm{km}\,\rm{s}^{-1}\,\rm{pc}^2$.
Based on $L'_{\rm{CO(4-3)}}$ and the scaling factors provided in Table\,2 (QSO) of \citet{Carilli2013} we estimate $L'_{\rm{CO(1-0)}} = L'_{\rm{CO(4-3)}}/0.87 = 2.32\times10^{10}\,\rm{K}\,\rm{km}\,\rm{s}^{-1}\,\rm{pc}^2$. 
We use $L'_{\rm{CO(1-0)}}$ to estimate the gas mass of J0341${+}$1720. Using the  $\alpha$ conversion factor for ultra luminous infrared galaxies (ULIRG, $L_{\rm{IR}}\geq10^{12}\,L_{\odot}$, see below), $\alpha_{\rm{CO}}\sim0.8\,M_{\odot} (\rm{K}\,\rm{km}\,\rm{s}^{-1}\,\rm{pc}^2)^{-1}$ \citep{Downes1998}, results in a gas mass of $M_{\rm{gas}}\approx 1.9\times10^{10}\,\rm{M}_\odot$.

In the following we will derive estimates for the dynamical mass of this system. The presented observations do not resolve the host galaxy emission and we rely on mean properties of other high-redshift quasar samples for this calculation.

We measured the FWHM of the CO(4-3) line from the Gaussian fit to the spectrum, $\rm{FWHM}_{\rm{CO(4-3)}}=460\pm69\,\rm{km}\,\rm{s}^{-1}$ ($\sigma_v=195\pm29\,\rm{km}\,\rm{s}^{-1}$). Based on the FWHM measurement we continue to assess the dynamical mass of the host galaxy by using the virial theorem for the case of a dispersion dominated system:
\begin{equation}
 M_{\rm{dyn,disp}} = \frac{3}{2} \frac{R_{\rm{CO}} \sigma_v^2}{G}
\end{equation}
where $G$ is the gravitational constant and $R_{\rm{CO}}$ is the radius of the line emitting region. As our mm observations do not spatially resolve the quasar host galaxy, we adopt a radius of $R_{\rm{CO}}=2.2\,\rm{kpc}$ for the source of the CO(4-3) emission, twice the effective (half-light) radius of \cii{} emission in high redshift quasar hosts (Neeleman et al. 2020, subm.; Novak et al. 2020, subm.) with a sample uncertainty of $\sigma=0.22\,\rm{kpc}$. Assuming that we can infer the gas velocity dispersion from the FWHM of the CO(4-3) we estimate a dispersion dominated dynamical mass of $M_{\rm{dyn,disp}} \approx 3\times10^{10}\,\rm{M}_\odot$.
If the system was rotationally supported with an inclination $i$, the dynamical mass can be approximated by \citep[see][]{Wang2013, Willott2015, Decarli2018}:
\begin{equation}
 M_{\rm{dyn,rot}} = \frac{R_{\rm{CO}}}{G} \cdot \left( \frac{3}{4} \frac{{\rm{FWHM}}}{\sin(i)}  \right)^2
\end{equation}
For J0341${+}$1720 we adopt an inclination of $i=33^{\circ}$, the mean inclination  for quasar hosts given in Neeleman et al. (2020, subm.). 
Using this assumption we infer a rotational dynamical mass of $M_{\rm{dyn,rot}} \approx 2\times10^{11}\,\rm{M}_\odot$.

Lastly, we use the continuum flux to estimate the star formation rate for J0341${+}$1720. Based on the observations in both, the lower and upper side-band, we measure an average continuum flux of $S_{\rm{obs, cont}}= 0.104(\pm0.013)\,\rm{mJy}$ at a frequency of $93.5\,\rm{GHz}$. This frequency probes the thermal infrared continuum of the quasar host. Assuming a modified blackbody with a dust temperature of $T_{\rm{dust}}=47\,\rm{K}$ and a power law slope $\beta=1.6$ \citep[see][their Equation\,2]{Beelen2006} we integrate the modified blackbody from $8$ to $1100\,\mu\rm{m}$ to estimate the total infrared luminosity of the continuum. 
We calculate a value of $L_{\rm{TIR}}\approx 1.0\times10^{13}\,\rm{L}_\odot$ making the host galaxy of J0341${+}$1720 a ULIRG.

In high redshift quasars the far-infrared emission stems from dust, which is predominantly heated by  stars \citep[e.g.,][]{Leipski2014, Venemans2017c}. Under this assumption we convert the TIR luminosity into a star formation rate of $\rm{SFR} \approx 1500\,\rm{M}_\odot\,\rm{yr}^{-1}$ \citep{Murphy2011}. 
By assuming that all dust emission is attributed to star formation, the derived star formation rate should be regarded as an upper limit. 
We caution that our data does not constrain the peak or shape of the SED and provide only a normalization of the chosen SED template. Thus our assumptions introduce large uncertainties of up to a factor of $\sim3$, as discussed in detail in \citet[][their Section\,4.1]{Venemans2018}.


\section{Discussion}\label{sec:discussion}

\begin{figure*}[t]
    \centering
    \includegraphics[width=\textwidth]{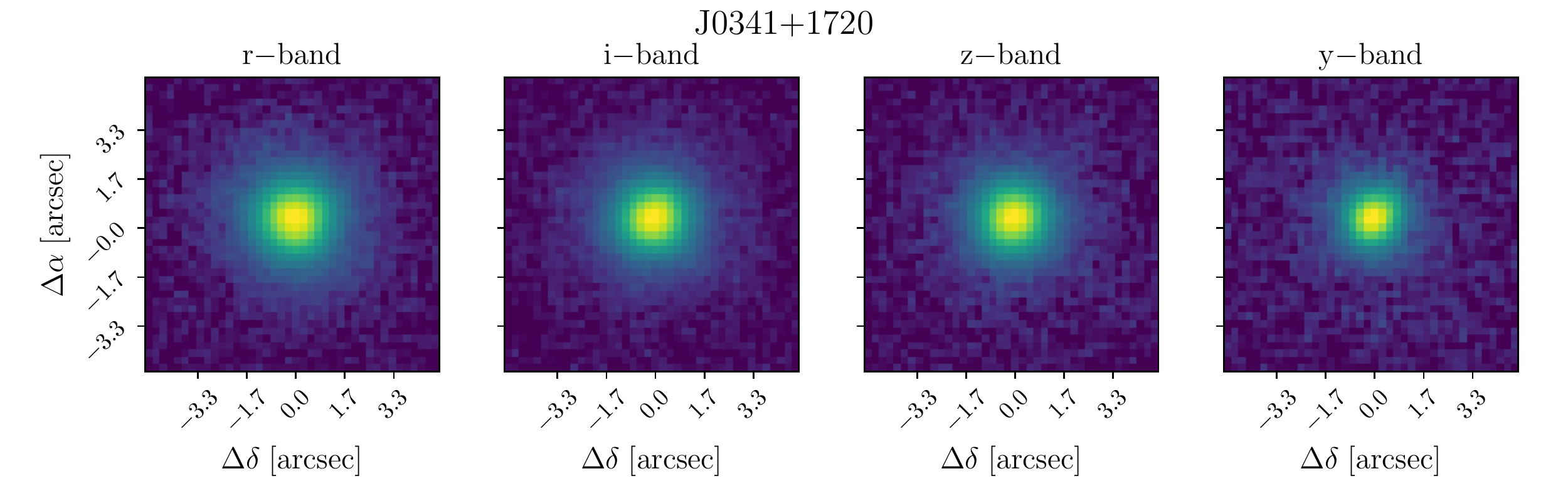}
    \includegraphics[width=\textwidth]{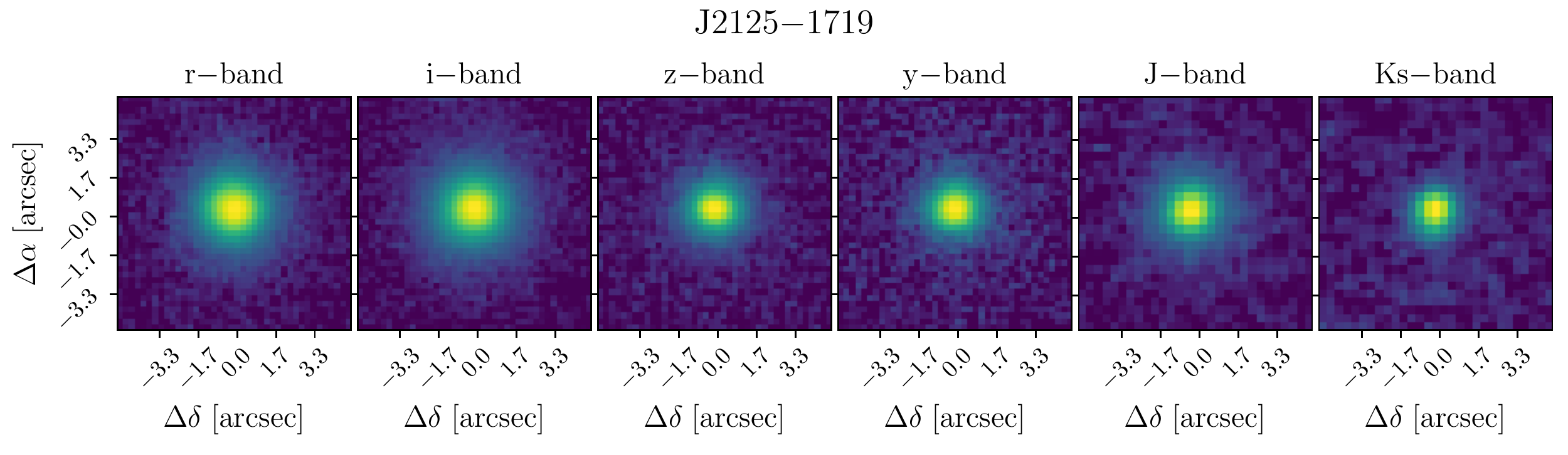}
    \caption{$10\arcsec{\times}10\arcsec$ cutout images of PS1 ($r$,$i$,$z$,$y$) photometry for J0341${+}$1720 and J2125${-}$1719 centered on the quasar position. We further included the VHS J- and Ks-band for J2125${-}$1719. Both quasars are well detected in all bands, showing a point-source-like PSF with no indication for multiple images or distortions as would be expected for strongly lensed systems.}
    \label{fig:photometry}
\end{figure*}

\subsection{Lensing}
Strong gravitational lensing can magnify the quasar's emission, which in turn leads to over-estimates on the measured luminosity, Eddington luminosity ratio and BH mass. As both J0341${+}$1720 and J2125${-}$1719 are extremely bright compared to the quasar population at the same redshifts (see Figure\,\ref{fig:zM1450}), the question arises whether their emission could have been amplified by a foreground galaxy (or galaxy cluster).

Figure\,\ref{fig:photometry} displays $10\arcsec{\times}10\arcsec$ cutout images of J0341${+}$1720 and J2125${-}$1719 from PS1 and the Vista Hemisphere Survey \citep[VHS,][]{McMahon2013}, where available. 
We measured the seeing by fitting a 2D Gaussian to the quasar and adjacent stars in a $400\times400\,\arcsec^2$ field of view. Our results are displayed in Table\,\ref{table:photometry} and show that both sources do not deviate from point-sources down to the seeing limit of the images.



\begin{deluxetable*}{c|cc|cc}
\tablecaption{Seeing measurements (FWHM of a 2D Gaussian) for PS1 and VHS photometry \label{table:photometry}}
\tablehead{\colhead{Photometric band} &\colhead{J0341${+}$1720} & \colhead{point sources} &\colhead{J2125${-}$1719} & \colhead{point sources}\\
\nocolhead{} & \colhead{FWHM (arcsec)} & \colhead{FWHM (arcsec)} & \colhead{FWHM (arcsec)} & \colhead{FWHM (arcsec)}}
\startdata 
r-band & $1.08$ & $1.07\pm0.06$ & $1.27$ & $1.23\pm0.05$\\
i-band & $0.96$ & $0.97\pm0.05$ & $1.13$ & $1.15\pm0.04$\\
z-band & $1.00$ & $0.99\pm0.04$ & $1.02$ & $1.03\pm0.06$\\
y-band & $0.88$ & $0.90\pm0.03$ & $1.13$ & $1.10\pm0.02$\\
\tableline
J-band & \dots & \dots & $0.95$ & $0.96\pm0.03$\\
K-band & \dots & \dots & $0.83$ & $0.82\pm0.03$\\
\enddata 
\end{deluxetable*}

A cross-match between the CASTLES catalog of lensed quasars and the Gaia DR2 source catalog \citep{GaiaCollab2016, GaiaCollab2018b} showed that Gaia can detect quasar pairs and quadruples down to a separation of $0\farcs5$.
A match of J0341${+}$1720 to Gaia DR2 finds that it is the only source listed within $20\arcsec$ of its position. In the case of J2125${-}$1719 a faint ($G=20.96$) second source is detected within $20\arcsec$ at a separation of $6\farcs43$. This separation would be too large to allow for substantial magnification of the quasar light. 
However, using Gaia we would not be able to identify lensed quasars with small image separations, such as J043947.08${+}$163415.7 at z=6.51 \citep{Fan2019} with an image separation of $\theta\sim0.2\arcsec$.
%

Both photometry and Gaia cross-matches suggest that neither J0341${+}$1720 or J2125${-}$1719 are being strongly lensed on scales of $\gtrsim1\farcs0$. 
Only high-resolution imaging using the Hubble Space Telescope or the Atacama Large Millimeter/submillimeter Array will be able to constrain lensing on smaller scales. For the remainder of this paper, we will assume these quasars not to be lensed.

\subsection{Black hole masses and super-Eddington accretion}

\begin{figure}
    \centering
    \includegraphics[width=0.45\textwidth]{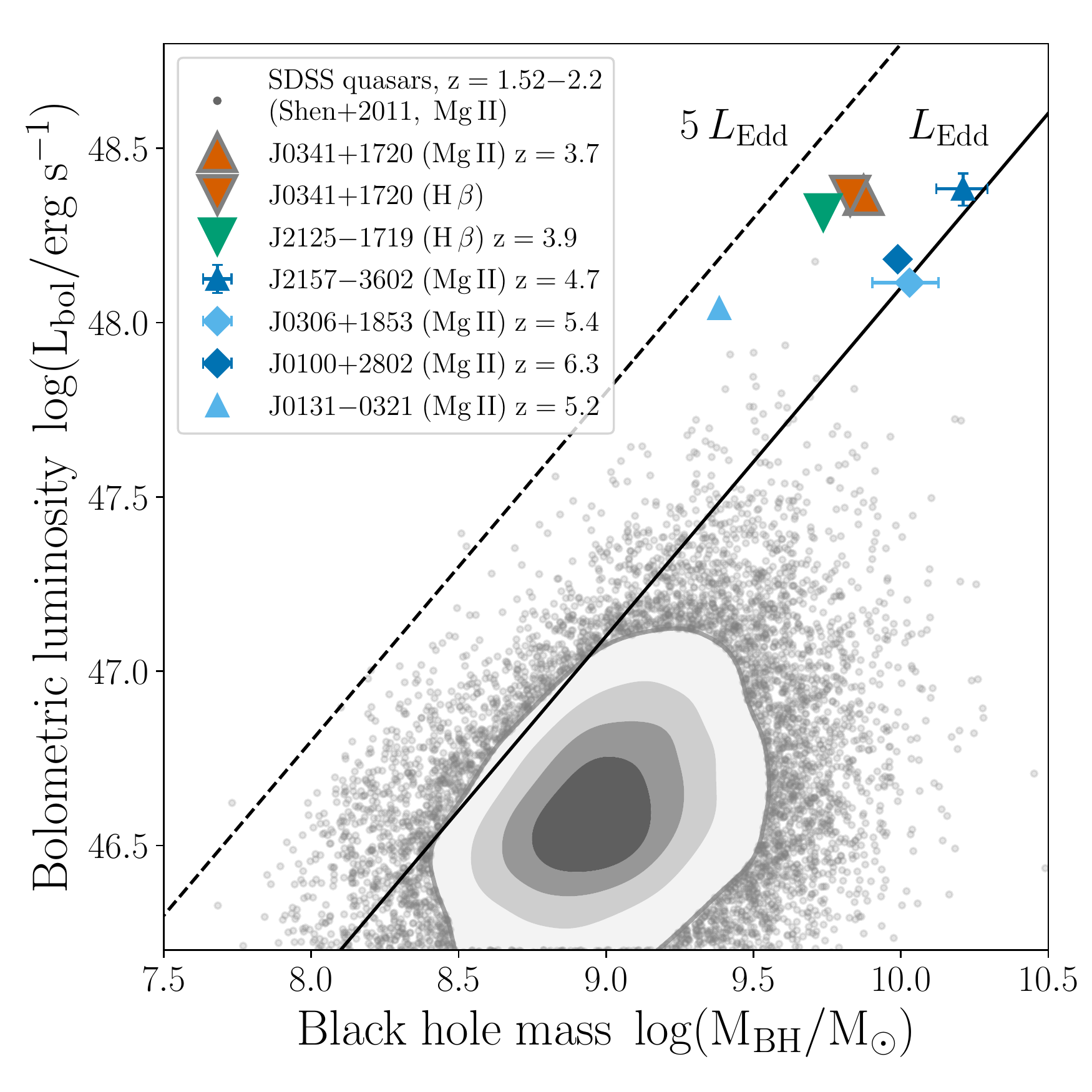}
    \caption{Bolometric luminosity as a function of black hole mass.
    We display the \ion{Mg}{2} and H$\beta$ measurements of J0341${+}$1720 and the H$\beta$ BH mass measurement for J2125${-}$1719. We compare them with the individual measurements for J2157{-}3602 \citep[][]{Onken2020}, SDSS\,J0131${-}$0321 \citep[][]{WeiMinYi2014}, SDSS\,J0306${+}$1853 \citep[][]{WangFeige2015}, and SDSS\,J0100+2802 \citep[as updated in][]{Schindler2020}. The symbols for each object are illustrated in the legend.
    We display the distribution of low-redshift ($z=1.52-2.2$) BH mass measurements of the SDSS quasar sample \citep{Shen2011} with filled grey contours. 
    We denote which emission line the BH mass estimate was based on in the legend and for consistency use the same single-epoch virial mass estimator for the same line. Where error bars are included, they show the statistical uncertainties of the measurements, while systematic uncertainties on the SMBH mass can be as large as $0.55\,\rm{dex}$ \citep{Vestergaard2009}.}
    \label{fig:bhmass_lbol}
\end{figure}

Figure\,\ref{fig:bhmass_lbol} shows the bolometric luminosities and black hole masses of J0341${+}$1720 (orange) and J2125${-}$1719 (green) in comparison with other measurements from the literature. 
The population of low-redshift SDSS quasars ($z{=}1.52{-}2.2$) with BH mass estimates based on detection of the \ion{Mg}{2} line is shown in grey contours \citep{Shen2011}.
Additionally, we add three ultra-luminous ($M_{1450}\leq-29$) quasars in the literature, SMSS\,J2157${-}$3602 \citep{Wolf2018, Onken2020} at $z\approx4.7$, SDSS\,J0306${+}$1853 \citep{WangFeige2015} at $z\approx5.4$, and SDSS\,J0100${+}$2802 \citep{Wu2015} at $z\approx6.3$, and an extremely luminous, super-Eddington quasar SDSS\,J0131${-}$0321 at $z=5.18$ \citep{WeiMinYi2014}. 
In all cases we adopt \ion{Mg}{2}-based BH mass measurements based on the relation of \citet{Vestergaard2009} for a valid comparison. In the case of SMSS\,J2157-3602 and SDSS\,J0131${-}$0321 we re-calculate their BH mass using their \ion{Mg}{2} FWHM and $L_{3000}$ and for SDSS\,J0100+2802 we use the value of \citet{Schindler2020}.
J0341${+}$1720 and J2125${-}$1719 are much more luminous compared to the population of low- and mid-redshift SDSS quasars. They harbor SMBHs with masses of $\log(M_{\rm{BH}}/M_{\odot})=9.5-10$ at the massive end of the low-redshift SDSS quasar distribution. 
While both quasars are similarly luminous to their three ultra-luminous siblings, they have lower BH masses, resulting in accretion rates moderately above the Eddington limit with $\lambda_{\rm{Edd}}=2-3$. 
The BH mass estimates from the H$\beta$ and \ion{Mg}{2} lines are generally regarded as robust. Do our measurements then provide tangible evidence of super-Eddington growth in these two systems?

In order to quantify some of the systematics associated with adopting a single-epoch virial estimator we adopt additional relations for the H$\beta$ and \ion{Mg}{2} lines to calculate the black hole masses.
In the case of J0341${+}$1720 we find the BH masses range between $M_{\rm{BH}}=6.73-15.55\times10^9\,M_{\odot}$ based on both the H$\beta$ and \ion{Mg}{2} using four different single-epoch virial mass estimators \citep{McLure2002, Vestergaard2006, Vestergaard2009, Shen2011}, resulting in Eddington luminosity ratios of $\lambda_{\rm{Edd}}{=}1.18{-}2.74$.
For J2125${-}$1719 we find BH masses of $M_{\rm{BH}}{=}4.53{-}7.53\times10^9\,M_{\odot}$ based on the H$\beta$ and \ion{C}{4} line using three different single-epoch virial mass estimators \citep{McLure2002, Vestergaard2006}, resulting in Eddington luminosity ratios of $\lambda_{\rm{Edd}}{=}2.18{-}3.62$.
Once we take the systematic uncertainty on the BH mass estimates of $0.55\,\rm{dex}$ into account and consider the range of BH masses derived above,  J0341${+}$1720 and J2125${-}$1719 could have accretion rates consistent with the Eddington limit.
In turn, they would also have BH masses on the order of $10^{10}\,M_{\odot}$.

\subsection{Black hole galaxy co-evolution}
Well established correlations between the masses of SMBHs and their host galaxy's bulge mass \citep[e.g.,][]{Magorrian1998, Marconi2003, Haering2004, Kormendy2013} suggest a coordinated co-evolution by a common physical mechanism \citep[e.g., see][]{Silk1998, DiMatteo2005, Hopkins2006, Angles2013, Peng2007, Jahnke2011}.
Active galactic nuclei and quasars have been used to investigate this correlation up to $z\sim7$ \citep[e.g.,][]{Walter2004, WangRan2010, Targett2012, Willott2015, Venemans2016, Izumi2019, Nguyen2020}.
In these cases the dynamical mass estimates from mm observations (see Section\,\ref{sec:J0341_mm}) have been used as an upper-limit proxy for the galaxy bulge mass. 

\begin{figure}
    \centering
    \includegraphics[width=0.45\textwidth]{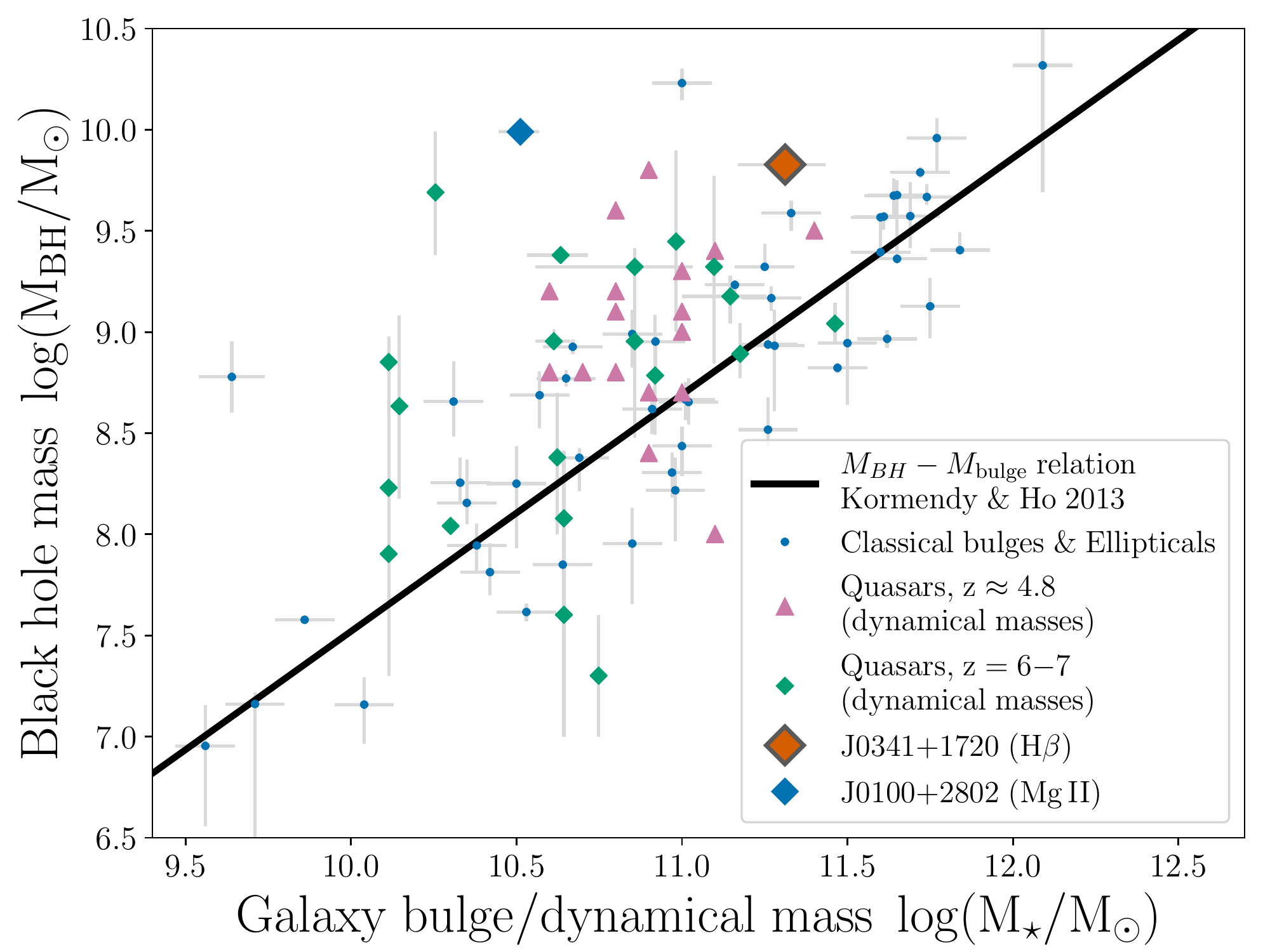}
    \caption{Black hole masses as a function of host galaxy masses. The orange diamond shows the measurement of the BH mass and galaxy host's dynamical mass (rotationally supported) for J0341${+}$1720.
    We include SDSS\,J0100+2802 as the blue diamond \citep{WangFeige2019a}. Purple triangles \citep{Nguyen2020} and green diamonds \citep{DeRosa2014, Willott2015, Venemans2016, Venemans2017b, Venemans2017c, Banados2018, Izumi2019, Onoue2019} show measurements from high-redshift quasar samples. Bulge mass measurements from local galaxies (blue dots) are taken from \citet{Kormendy2013}. 
    Similar to $z\gtrsim4.8$ quasars J0341${+}$1720 lies off-set above from the local relation.
    }
    \label{fig:coevolution}
\end{figure}

In Figure\,\ref{fig:coevolution} we put our measurements of the BH mass and the host galaxy dynamical (rotationally supported) mass of J0341${+}$1720 in context with measurements in the literature. 
The local black hole mass galaxy bulge mass relation is shown in black along with individual measurements from classical bulges and elliptical galaxies (blue dots) \citep{Kormendy2013}.
With purple triangles and green diamonds we further display measurements from quasars at $z=4.8$ \citep{Nguyen2020} and at $z=6-7$ \citep{DeRosa2014, Willott2015, Venemans2016, Venemans2017b, Venemans2017c, Banados2018, Izumi2019, Onoue2019}, respectively.
The majority of $z{\approx}6{-}7$ redshift quasars are found above the local relation with the exception of a few low-luminosity high-redshift quasars \citep{Izumi2019, Onoue2019}, which scatter below. SDSS\,J0100+2802 \citep[][blue diamond]{WangFeige2019a} is a prominent example of an ultra-luminous, high-redshift quasar lying well above the local relation.
J0341${+}$1720 is highlighted with an orange diamond above the local relation. Based on our dynamical mass estimate (rotationally supported) the host galaxy is only $\sim30$ times more massive than the quasar. Assuming that the quasar and the galaxy will grow continuously over the next $10\,\rm{Myr}$, we find that this system is moving even further away from the local relation.
However, we need to stress that our dynamical mass estimate is based on unresolved mm observations and several assumptions were made in the calculation (see Section\,\ref{sec:J0341_mm}). For better constraints on the host galaxy properties resolved data will be necessary.

\section{Summary}\label{sec:conclusion}

In this work we have taken a closer look at two ultra-luminous quasars, J0341${+}$1720 and J2125${-}$1719. Analysis of their full rest-frame UV to optical spectra revealed SMBHs with masses of $M_{\rm{BH}}=6.73\times10^{9}\,M_{\odot}$ and $M_{\rm{BH}}=5.45\times10^{9}\,M_{\odot}$, resulting in Eddington luminosity ratios of $\lambda_{\rm{Edd}}=2.74$ and $\lambda_{\rm{Edd}}=3.01$. 
Their SMBHs are among the most massive compared to black holes known (see Figure\,\ref{fig:bhmass_lbol}) and are rapidly accreting new material, possibly beyond the Eddington limit.
We further observed the host galaxy emission of J0341${+}$1720 at mm wavelengths with a clear detection of the CO(4-3) transition and the underlying continuum.
We estimate dispersion dominated and rotationally supported dynamical masses of $M_{\rm{dyn,disp}} \approx 3\times10^{10}\,\rm{M}_\odot$ and $M_{\rm{dyn,rot}}\approx2\times10^{11}\,M_{\odot}$, respectively. Similarly to quasars at $z=5-7$, J0341${+}$1720 lies above the local SMBH galaxy scaling relations (see Figure\,\ref{fig:coevolution}).
Based on its total infrared luminosity ($L_{\rm{TIR}}\approx1.0\times10^{13}\,L_{\odot}$) the host galaxy of J0341${+}$1720 can be classified as a ULIRG with an approximate star formation rate of $\rm{SFR}\approx1500\,M_{\odot}\,\rm{yr}^{-1}$. Given these estimates the system would evolve even further away from the local relation.

The rapid assembly of billion solar mass SMBHs in the early universe \citep{Banados2018, Onoue2020, YangJinyi2020} poses challenges to standard scenarios of SMBH formation and evolution. Bounded by the Eddington limit, they could not grown to their current masses in the time since the Big Bang unless their seed masses were very high ($M_{\rm{seed}} > 10^4 M_{\odot}$). 
The presented analysis, highlighting two systems with evidence for super-Eddington accretion, can help to resolve some of this tension.
In their recent review \citet{Inayoshi2019_arxiv} point out that the time-averaged mass accretion rate in the most massive and highest-redshift SMBHs only needs to be moderately above the Eddington limit to explain their observed masses within most BH seeding models.
However, it remains an open question how such high accretion rates can be sustained, considering that the host galaxy of J0341${+}$1720 is only ${\sim}30$ more massive than its SMBH. Only resolved mm observations will be able to unveil the mass, dynamics and extent of the large gas reservoir fueling the quasars' emission.

\appendix 

\section{Modelling of the optical and near-infrared spectra}\label{app:fitmethod}

We have used a custom interactive fitting code based on the LMFIT python package \citep{lmfit2014} package to model the spectra of both quasars. This is a two-stage process, in which we interactively set the continuum and line emission regions, add models for the continuum and the emission lines and determine the initial best fit, which is then saved. In a second step we re-sample each spectrum 1000 times by randomly drawing new flux values on a pixel by pixel bases from a Gaussian distribution set by the original flux values and their uncertainties. All re-sampled spectra are then fit using our interactively determined best-fit as the initial guess. 
We build posterior distributions for all fit parameters by recording the best-fit value from each re-sampled spectral fit. The results presented here refer to the median of this distribution and the associated uncertainties are the $13.6$ and $86.4$ percentile values. 

The spectral fits constructed with our interactive code consist of continuum and line models. All continuum models are subtracted from the spectrum before the line models are fit. We will now discuss the components of our continuum model and provide general properties of the emission lines included in our fits. 

\subsection{The continuum model}

We model the quasar continuum with three general components. First we approximate the non-thermal radiation from the accretion disk by a single power-law normalized at $2500\,\text{\AA}$:
\begin{equation}
    F_{\rm{PL}}(\lambda) = F_{\rm{PL}, 0}\ \left(\frac{\lambda}{2500\,\text{\AA}}\right)^{\alpha_{\lambda}} \ .
\end{equation}
Here $F_{\rm{PL}, 0}$ is the normalization and $\alpha_{\lambda}$ is the slope of the power law.

In addition to the power law, high-order Balmer lines and bound-free Balmer continuum emission give rise to a Balmer pseudo-continuum. We do not model the region, where the high-order Balmer lines merge and thus we only model the bound-free emission blue-ward of the Balmer break at $\lambda_{\rm{BE}} = 3646\,\text{\AA}$. 
The Balmer continuum models follows the description of \citet{Dietrich2003c}, who assumed the Balmer emission arises from gas clouds of uniform electron temperature that are partially optically thick:
\begin{equation}
\begin{split}
   F_{\rm{BC}}(\lambda) =  F_{\rm{BC}, 0}\ B_{\lambda}(\lambda, T_e) \left(1 - e^{\left(-\tau_{\rm{BE}} (\lambda/\lambda_{\rm{BE}})^3 \right)} \right) \ ,  \\
   \forall\ \lambda \le \lambda_{\rm{BE}} \ ,
 \end{split}
\end{equation}
where $B_{\lambda}(T_e)$ is the Planck function at the electron temperature of $T_e$, $\tau_{\rm{BE}}$ is the optical depth at the Balmer edge and $F_{\rm{BC}, 0}$ is the normalized flux density at the Balmer break \citep{Grandi1982}. 
We estimate the strength of the Balmer emission, $F_{\rm{BC}, 0}$, from the flux density slightly redward of the Balmer break at $\lambda = 3675\,\text{\AA}$ after subtraction of the power law continuum \citep{Dietrich2003c}. We further fix the electron temperature and the optical depth to values of $T_e = 15,000\,\rm{K}$  and $\tau_{\text{BE}}=1$, common values in the literature \citep{Dietrich2003c, Kurk2007, DeRosa2011, Mazzucchelli2017, Shin2019, Onoue2020}.

Many quasar spectra show a strong contribution from transitions of single and double ionized iron atoms (\ion{Fe}{2} and \ion{Fe}{3}), which are important to correctly model the broad \ion{Mg}{2} line as well as the H$\beta$ and O[III] lines.

The large number of iron transitions, especially from \ion{Fe}{2}, lead to a multitude of emission lines, which blend into an iron pseudo-continuum. We adopt the empirical and semi-empirical iron templates, derived from the narrow-line Seyfert 1 galaxy I Zwicky 1 \citep{Boroson1992a, Vestergaard2001, Tsuzuki2006} to model the iron emission for the two quasar spectra.

To accurately model the \ion{Mg}{2} inclusion of the surrounding iron emission is crucial ($2200-3500\,\text{\AA}$). As discussed in \citet{Onoue2020} and \citet{Schindler2020} the \citet{Tsuzuki2006} template, which includes an iron contribution beneath the \ion{Mg}{2} is preferable over the \citet{Vestergaard2001} in this region when measuring the \ion{Fe}{2} or \ion{Mg}{2} properties. However, in order to use the black hole mass scaling relations for \ion{Mg}{2}, which were established with FWHM measurements using the \citep{Vestergaard2001} template, we fit J0341${+}$1720 with each template. 

We aim to measure the properties of the H$\beta$ line for both quasars. Similar to \ion{Mg}{2} this line is emitted in a region with significant \ion{Fe}{2} contribution. We adopt the empirical iron template of \citet{Boroson1992a} in this region ($3700-5600\,\text{\AA}$).

The iron templates are redshifted to the systemic redshift of the quasars. In addition, we convolve the templates with a Gaussian kernel to broaden the intrinsic width of the iron emission 
of I Zwicky 1, $ \rm{FWHM} \approx 900\,\text{km}\text{s}^{-1}$, according to the quasar's broad lines \citep[see][for a discussion]{Boroson1992a}:
\begin{equation}
 \sigma_{\rm{conv}} = \sqrt{\rm{FWHM}_{\rm{obs}}^2  -\rm{FWHM}_{\rm{I\ Zwicky\ 1}}^2}/ 2\sqrt{2\ln2}
\end{equation}
We set the FWHM and the redshift of the iron template in the \ion{Mg}{2} region to the values determined from the \ion{Mg}{2} line, while the H$\beta$ redshift and FWHM are applied to the iron template at $3700-5600\,\text{\AA}$ \citep[see ][for a similar approach]{Tsuzuki2006, Shin2019}.
The full continuum model, including the broadened iron template, and the emission line models are fit iteratively until the FWHM of the \ion{Mg}{2} and H$\beta$ line converge. 

Our sources are extremely luminous quasars and we therefore do not include a contribution from the stellar component of the quasar host. 

We fit these three components of our continuum model to the spectra in line-free regions. Which regions in a quasar spectrum can be considered line-free is widely discussed in the literature \citep[e.g.,][]{Vestergaard2006, Decarli2010, Shen2011, Mazzucchelli2017, Shen2019b}. We follow \citet{Vestergaard2006} and \citet{Shen2011} and adopt the following regions in our continuum fit:
$\lambda_{\rm{rest}} = 1265{-}1290\,\text{\AA}$, $1340{-}1375\,\text{\AA}$, $1425{-}1470\,\text{\AA}$, $1680{-}1705\,\text{\AA}$,  $2200{-}2700\,\text{\AA}$, $4435{-}4700\,\text{\AA}$.
Unfortunately, the spectral coverage makes it impossible to include regions red-ward of the \ion{Mg}{2} line in the fit of J0341${+}$1720 and red-ward of H$\beta$ in both fits. We interactively adjust the continuum windows to exclude regions with absorption lines, sky-line residuals or unusually large flux errors. The specific regions included in the continuum fit are shown in Figures\,\ref{fig:specJ0341} and \ref{fig:specJ2125} as the light blue regions on the top of each panel. 

\subsection{Emission line models}

We focus our analysis of the optical and near-infrared quasar spectra on the broad \ion{Si}{4}, \ion{C}{4}, \ion{Mg}{2}, H$\beta$ and O[III] lines. Given the signal-to-noise ratio and low to medium resolution of the spectra, we do not sufficiently resolve any of emission line doublets or triplets and therefore model them as single lines with rest-frame wavelengths of  $\lambda1399.8\,\text{\AA}$ for \ion{Si}{4}, $\lambda1549.06\,\text{\AA}$ for \ion{C}{4}, $\lambda2798.75\,\text{\AA}$ for \ion{Mg}{2}, $\lambda4862.68\,\text{\AA}$ for H$\beta$, and $\lambda4960.30\,\text{\AA}$/$\lambda5008.24\,\text{\AA}$ for the two [O~III] lines \citep[see][]{VandenBerk2001}. 
The lines are modeled with one or two Gaussian profiles, depending on the line shape. 

The broad \ion{Si}{4} line blends together with the close-by semi-forbidden O~IV] $\lambda1402.06\,\text{\AA}$ transition. Given the resolution and the quality of our data we cannot disentangle the two lines and rather model their blend, \ion{Si}{4}+O~IV] $\lambda1399.8\,\text{\AA}$\footnote{http://classic.sdss.org/dr6/algorithms/linestable.html}.


\acknowledgements{
Mladen Novak, Bram Venemans, and Fabian Walter acknowledge support from the ERC Advanced Grant 740246 (Cosmic Gas).

Feige Wang thanks the supports provided by NASA through the NASA Hubble Fellowship grant \#HST-HF2-51448.001-A awarded by the Space Telescope Science Institute, which is operated by the Association of Universities for Research in Astronomy, Incorporated, under NASA contract NAS5-26555.

This work is based on observations carried out under project number S19CZ with the IRAM NOEMA Interferometer [30m telescope]. IRAM is supported by INSU/CNRS (France), MPG (Germany) and IGN (Spain). The authors would like to than Jan-Martin Winters for his support with the calibration and analysis of the interferometric data.

Based on observations with the VATT: the Alice P. Lennon Telescope and the Thomas J. Bannan Astrophysics Facility

Based on observations obtained as part of the VISTA Hemisphere Survey, ESO Progam, 179.A-2010 (PI: McMahon)

Funding for the Sloan Digital Sky Survey IV has been provided by the Alfred P. Sloan Foundation, the U.S. Department of Energy Office of Science, and the Participating Institutions. SDSS-IV acknowledges
support and resources from the Center for High-Performance Computing at
the University of Utah. The SDSS web site is www.sdss.org.
SDSS-IV is managed by the Astrophysical Research Consortium for the 
Participating Institutions of the SDSS Collaboration including the 
Brazilian Participation Group, the Carnegie Institution for Science, 
Carnegie Mellon University, the Chilean Participation Group, the French Participation Group, Harvard-Smithsonian Center for Astrophysics, 
Instituto de Astrof\'isica de Canarias, The Johns Hopkins University, Kavli Institute for the Physics and Mathematics of the Universe (IPMU) / 
University of Tokyo, the Korean Participation Group, Lawrence Berkeley National Laboratory, 
Leibniz Institut f\"ur Astrophysik Potsdam (AIP),  
Max-Planck-Institut f\"ur Astronomie (MPIA Heidelberg), 
Max-Planck-Institut f\"ur Astrophysik (MPA Garching), 
Max-Planck-Institut f\"ur Extraterrestrische Physik (MPE), 
National Astronomical Observatories of China, New Mexico State University, 
New York University, University of Notre Dame, 
Observat\'ario Nacional / MCTI, The Ohio State University, 
Pennsylvania State University, Shanghai Astronomical Observatory, 
United Kingdom Participation Group,
Universidad Nacional Aut\'onoma de M\'exico, University of Arizona, 
University of Colorado Boulder, University of Oxford, University of Portsmouth, 
University of Utah, University of Virginia, University of Washington, University of Wisconsin, 
Vanderbilt University, and Yale University.

This work has made use of data from the European Space Agency (ESA) mission
{\it Gaia} (\url{https://www.cosmos.esa.int/gaia}), processed by the {\it Gaia}
Data Processing and Analysis Consortium (DPAC,
\url{https://www.cosmos.esa.int/web/gaia/dpac/consortium}). Funding for the DPAC
has been provided by national institutions, in particular the institutions
participating in the {\it Gaia} Multilateral Agreement.
}

\facilities{LBT (LUCI1, LUCI2), Magellan:Baade (FIRE), VATT (VATTSpec), SOAR (Goodman High Throughput Spectrograph), IRAM:NOEMA}

\software
{Astropy \citep{astropy1, astropy2}, SciPy \citep{scipy}, Numpy \citep{numpy}, Pandas \citep{pandas_software, pandas_paper}, LMFIT \citep{lmfit2014}, Pypeit \citep{PypeitProchaska2019, PypeitProchaska2020}, Extinction \citep{python_extinction}
}

\bibliography{all}{}
\bibliographystyle{aasjournal}

\end{document}

%% file: prop_table.tex
\begin{deluxetable}{ccc}
\tabletypesize{\footnotesize} 
\tablecaption{Quasar properties directly measured or derived from the spectral fits \label{tab:fitproperties}}
\tablehead{\colhead{Measured Property} &\colhead{J0341+1720} &\colhead{J2125-1719} } 
\startdata 
$M_{1450}/(\rm{mag})$ &${-29.562}_{+0.004}^{-0.004}$ & ${-29.386}_{+0.001}^{-0.001}$\\ 
$L_{\rm{3000}} /(10^{46}\,\rm{erg}\,\rm{s}^{-1})$ & ${44.96}_{-0.21}^{+0.22}$ & ${40.13}_{-0.10}^{+0.11}$\\ 
$L_{\rm{3000}, VW01} /(10^{46}\,\rm{erg}\,\rm{s}^{-1})$ & ${44.98}_{-0.21}^{+0.20}$ &  \dots\\ 
$L_{\rm{5100}} /(10^{46}\,\rm{erg}\,\rm{s}^{-1})$ & ${27.22}_{-0.25}^{+0.25}$ & ${25.26}_{-0.13}^{+0.14}$\\ 
Power law index $\alpha$ & ${-1.620}_{-0.008}^{+0.007}$ & ${-1.551}_{-0.004}^{+0.004}$\\ 
$F_{\rm{FeII}}/F_{\rm{MgII}}$ & ${2.78}_{-0.65}^{+0.79}$ & \dots\\ 
\tableline Derived property & J0341+1720 & J2125-1719 \\ 
\tableline $L_{\rm{bol}} /(10^{48}\,\rm{erg}\,\rm{s}^{-1})$ & ${2.32}_{-0.01}^{+0.01}$ & ${2.07}_{-0.01}^{+0.01}$\\ 
MgII BH mass (VW09)/($10^{9}\,\rm{M}_{\odot}$) & ${7.47}_{-1.35}^{+1.63}$ & \dots\\ 
$\lambda_{\rm{Edd, MgII}}$ (VW09) & ${2.47}_{+0.54}^{-0.44}$ & \dots\\ 
H$\beta$ BH mass (VO06) ($10^{9}\,\rm{M}_{\odot}$) & ${6.73}_{-0.83}^{+0.75}$ & ${5.45}_{-0.55}^{+0.60}$\\ 
$\lambda_{\rm{Edd, H}\beta}$  (VO06) & ${2.74}_{+0.39}^{-0.27}$ & ${3.01}_{+0.34}^{-0.30}$\\ 
\ion{C}{4} BH mass (VO06) ($10^{9}\,\rm{M}_{\odot}$) & ${33.80}_{-2.43}^{+2.11}$ & ${4.53}_{-0.38}^{+0.36}$\\ 
$\lambda_{\rm{Edd,CIV}}$  (VO06) & ${0.55}_{+0.04}^{-0.03}$ & ${3.63}_{+0.33}^{-0.27}$\\ 
\ion{C}{4} BH mass (Co17) ($10^{9}\,\rm{M}_{\odot}$) & ${14.20}_{-1.16}^{+1.22}$ & \dots\\ 
$\lambda_{\rm{Edd,CIV}}$  (Co17) & ${1.30}_{-0.10}^{+0.12}$ & \dots\\ 
\enddata 
\tablerefs{VW01 : Iron template of \citet{Vestergaard2001}; VW09 : BH mass estimator of \citet{Vestergaard2009}; VO06 : BH mass estimator of \citet{Vestergaard2006}; Co17 : BH mass estimator of \citet{Coatman2017}} 
\tablecomments{All properties of J0341{+}1720 are from the fit with the \citet{Tsuzuki2006} iron template with the exception of $L_{3000,\rm{VW01}}$, for which the \citet{Vestergaard2001} iron template was used.} 
\end{deluxetable} 

%% file: line_table.tex
\begin{deluxetable*}{ccccccc}
\tabletypesize{\footnotesize} 
\tablecaption{Emission line properties of J0341+1720 and J2125-1719 \label{tab:line_measurements}}
\tablehead{\colhead{Line} &\colhead{$z_{\rm{line, peak}}$} &\colhead{$F_{\rm{line}}$} &\colhead{FWHM$_{\rm{line}}$} &\colhead{EW$_{\rm{line}}$} &\colhead{$\Delta v({z_{\rm{line}}}{-}z_{\rm{sys}})$} & \colhead{Components}  \\ 
\nocolhead{} &\nocolhead{} &\colhead{($10^{-17}\,\rm{erg}\,\rm{s}^{-1}\rm{cm}^2$)} &\colhead{$(\rm{km}\,\rm{s}^{-1})$} &\colhead{(\AA)} &\colhead{$(\rm{km}\,\rm{s}^{-1})$} &\nocolhead{}} 
\startdata 
\multicolumn{7}{c}{J0341+1720} \\ 
\tableline\ion{Si}{4} &${3.6638}_{-0.0018}^{+0.0021}$ & ${2459.97}_{-134.59}^{+132.59}$ & ${7588}_{-329}^{+311}$ & ${7.11}_{-0.41}^{+0.41}$ & ${-2813}_{-114}^{+135}$ & 1G\\ 
\ion{C}{4} &${3.6691}_{-0.0013}^{+0.0013}$ & ${4419.30}_{-155.00}^{+127.42}$ & ${8484}_{-310}^{+261}$ & ${15.00}_{-0.58}^{+0.49}$ & ${-2473}_{-85}^{+85}$ & 1G\\ 
\ion{Mg}{2} &${3.7024}_{-0.0016}^{+0.0017}$ & ${783.69}_{-52.05}^{+59.00}$ & ${3551}_{-316}^{+362}$ & ${6.01}_{-0.40}^{+0.46}$ & ${-342}_{-105}^{+108}$ & 1G\\ 
\ion{Mg}{2} (VW01) &${3.7014}_{-0.0017}^{+0.0016}$ & ${929.43}_{-69.88}^{+68.58}$ & ${3922}_{-371}^{+407}$ & ${7.23}_{-0.55}^{+0.55}$ & ${-411}_{-111}^{+103}$ & 1G\\ 
$\rm{H}\beta$ &${3.7035}_{-0.0006}^{+0.0006}$ & ${2274.75}_{-70.91}^{+69.02}$ & ${3983}_{-255}^{+215}$ & ${45.90}_{-1.65}^{+1.59}$ & ${-275}_{-36}^{+38}$ & 2G\\ 
CO(4-3) & $3.7078\pm0.0005$ & \dots & $460\pm69$ & \dots & $0\pm69$ & 1G \\ 
\tableline\multicolumn{7}{c}{J2125-1719} \\ 
\tableline\ion{Si}{4} &${3.8886}_{-0.0003}^{+0.0003}$ & ${1615.57}_{-18.85}^{+17.39}$ & ${4193}_{-76}^{+42}$ & ${6.28}_{-0.08}^{+0.07}$ & ${-765}_{-17}^{+17}$ & 1G\\ 
\ion{C}{4} &${3.8983}_{-0.0019}^{+0.0026}$ & ${3887.95}_{-110.13}^{+177.84}$ & ${3246}_{-140}^{+126}$ & ${17.58}_{-0.50}^{+0.82}$ & ${-169}_{-117}^{+156}$ & 2G\\ 
$\rm{H}\beta$ &${3.9032}_{-0.0007}^{+0.0008}$ & ${1176.19}_{-28.00}^{+25.84}$ & ${3653}_{-189}^{+197}$ & ${31.58}_{-0.85}^{+0.73}$ & ${131}_{-42}^{+47}$ & 2G\\ 
$\rm{O[III]}\ \lambda4960\,$\AA &${3.9010}_{-0.0005}^{+0.0005}$ & ${139.26}_{-12.48}^{+13.36}$ & ${843}_{-95}^{+98}$ & ${3.81}_{-0.34}^{+0.37}$ & ${-6}_{-29}^{+32}$ & 1G\\ 
$\rm{O[III]}\ \lambda5007\,$\AA &${3.9011}_{-0.0003}^{+0.0003}$ & ${347.49}_{-16.36}^{+17.59}$ & ${811}_{-54}^{+47}$ & ${9.58}_{-0.45}^{+0.49}$ & ${-0}_{-17}^{+17}$ & 1G\\ 
\enddata 
\tablecomments{All line measurements of J0341{+}1720 are from the fit with the \citet{Tsuzuki2006} iron template with the exception of \ion{Mg}{2} (VW01), which was fit using the \citet{Vestergaard2001} iron template.} 
\end{deluxetable*} 